\documentclass[usenatbib]{mn2e}
\usepackage{aas_macro, amsmath, amssymb, color, graphicx, multirow, my_macro}
\defcitealias{Tinker2011}{Paper~I}
\defcitealias{Wetzel2012}{Paper~II}
\defcitealias{Wetzel2013}{Paper~III}
\defcitealias{Wetzel2013a}{Paper~IV}
\defcitealias{Wetzel2013b}{Paper~V}
\defcitealias{Tinker2013}{Tinker et al., in prep.}
\defcitealias{WetzelPrep}{Wetzel et al., in prep.}

\begin{document}  
\title[Galaxy evolution near groups and clusters]{Galaxy evolution near groups and clusters:
ejected satellites and the spatial extent of environmental quenching}
\author[Wetzel, Tinker, Conroy \& van den Bosch]{Andrew R. Wetzel${}^1$, Jeremy L. Tinker${}^2$, Charlie Conroy${}^3$, and Frank C. van den Bosch${}^1$\\
$^{1}$Department of Astronomy, Yale University, New Haven, CT 06520, USA\\
$^{2}$Center for Cosmology and Particle Physics, Department of Physics, New York University, New York, NY 10013, USA\\
$^{3}$Department of Astronomy \& Astrophysics, University of California, Santa Cruz, CA 95064, USA
}
\date{March 2013}
\pagerange{\pageref{firstpage}--\pageref{lastpage}} \pubyear{2013}
\maketitle
\label{firstpage}

%===================================================================================================
\begin{abstract}
Galaxies that are several virial radii beyond groups/clusters show preferentially quiescent star formation rates.
Using a galaxy group/cluster catalog from the Sloan Digital Sky Survey, together with a cosmological \textit{N}-body simulation, we examine the origin of this environmental quenching beyond the virial radius.
Accounting for the clustering of groups/clusters, we show that central galaxies show enhanced SFR quenching out to 2.5 virial radii beyond groups/clusters, and we demonstrate that this extended environmental enhancement can be explained simply by `ejected' satellite galaxies that orbit beyond their host halo's virial radius.
We show that ejected satellites typically orbit for several Gyr beyond the virial radius before falling back in, and thus they compose up to 40\% of all central galaxies near groups/clusters.
We show that a model in which ejected satellites experience the same SFR quenching as satellites within a host halo can explain essentially all environmental dependence of galaxy quenching.
Furthermore, ejected satellites (continue to) lose significant halo mass, an effect that is potentially observable via gravitational lensing.
The SFRs/colors and stellar-to-halo masses of ejected satellites highlight the importance of environmental \textit{history} and present challenges to models of galaxy occupation that ignore such history.
\end{abstract}


%===================================================================================================
\section{Introduction}

Environment impacts galaxy evolution.
Galaxies in groups and clusters are more likely to have suppressed (`quiescent') star formation rates (SFR) and lie on the red sequence, have more elliptical/bulge-dominated morphologies, and have less atomic/molecular gas than galaxies of similar stellar mass in comparatively isolated `field' environments \citep[][for review]{Oemler1974, Dressler1980, DresslerGunn1983, Balogh1997, BlantonMoustakas2009}.
While these differences are strongest for galaxies in massive clusters, they persist in groups at least down to virial masses of $\sim 10 ^ {12} \msun$ \citep[e.g.,][]{Weinmann2006a, Wetzel2012}, comparable to the Milky Way and M31, whose satellites also show such environmental trends \citep[][for review]{Mateo1998}.

Several observational works have demonstrated that this environmental dependence is confined mostly to physical scales corresponding to the virial radius of a group/cluster and is driven by satellite galaxies, which are not the massive `central' galaxy at the core of a host halo \citep[e.g.,][]{Hogg2004, Kauffmann2004, Weinmann2006a, BlantonBerlind2007, vandenBosch2008, Tinker2011, Wetzel2012}.
These results are physically motivated, given that a host halo's virial radius broadly corresponds to a physical transition from the low-density `field' environment to a high-density region where dark matter and gas are virialized.

While galaxies within groups/clusters strongly exhibit the above environmental dependencies, such trends can extend to galaxies beyond the virial radius of a group/cluster.
Several authors have noted an enhanced quiescent/red fraction and reduced $\hi$ emission in galaxies out to $2 - 4$ virial radii around massive clusters \citep{Balogh2000, Solanes2002, Hansen2009, vonderLinden2010, Wetzel2012}, as well as for dwarf galaxies around lower mass groups (Y. \citealt{Wang2009b, Geha2012}), including those near the Milky Way and M31 \citep{Fraternali2009}.

The above observational trends are plausibly driven, at least in part, by these galaxies having passed within much smaller distances from a group/cluster.
Indeed, \citet{Balogh2000} first noted from \textit{N}-body simulations of clusters that particles that have passed within the virial radius can then orbit well outside of it.
Satellite galaxies can do so as well, as investigated both through analytic arguments \citep{Mamon2004} and in numerical simulations \citep{Gill2005, Sales2007b, Warnick2008, Ludlow2009, Wang2009a, Knebe2011, Mahajan2011, Teyssier2012, SinhaHolleyBockelmann2012, Bahe2013}.
In general, these works found that these `ejected' (or `backsplash') satellites typically orbit back out to a maximum distance of $4\,\rthc \approx 2.5\,\rthm$\footnote{
We define a halo's virial radius, $\rdelta$, such that the average interior density is $\Delta$ times a reference density, $\rho_{\rm ref}$: $\mdelta = \Delta \frac{4}{3} \pi \rho_{\rm ref} \rdelta ^ 3$.
$\Delta = 200$m means $200 \times$ the average matter density of the Universe, and $\Delta = 200$c means $200 \times$ the critical density.
}
beyond a host halo after passing through it, and that as many as half of all galaxies within this distance are composed of these ejected satellites, with higher fractions for less massive galaxies and around more massive host halos.
Thus, ejected satellites are potentially critical for understanding the properties of galaxies near groups/clusters and obtaining a complete picture of environmental dependence.

Indeed, several works suggest that ejected satellites are preferentially quiescent.
Y. \citet{Wang2009b} used the SDSS group catalog of \citet{Yang2007} to examine the $g - r$ colors of faint ($M_r - 5 \log(h) > -17$) central galaxies.
They found that the reddest population lies within $\sim3\,R_{180\rm m}$ of a nearby massive host halo and therefore argued that such central galaxies are likely ejected satellites.
Relatedly, analyzing more massive galaxies near massive clusters also via the group catalog of \citet{Yang2007}, \citet{Mahajan2011} statistically compared the phase-space distribution (in redshift space) of galaxies near clusters with that of particles taken from a smooth particle hydrodynamics (SPH) simulation, arguing that ejected satellites are quenched significantly relative to infalling galaxies and that star formation quenching occurs after a single pericentric passage.
More dramatically, \citet{Geha2012} found that all spectroscopically quiescent galaxies with $\mstar < 10 ^ 9 \msun$ lie within $1.5 \mpc$ of a more massive galaxy, implying that \textit{all} low-mass quiescent galaxies are either satellites or ejected satellites.\footnote{
While Y. \citet{Wang2009b} found that more than half of their red galaxies do \textit{not} lie within $\sim 3\,R_{180\rm m}$ of a more massive halo, the results of \citet{Geha2012}, using spectroscopic measures of star formation, imply that the population of isolated, red galaxies in \citeauthor{Wang2009b} arises from dust reddening---a significant effect for low-mass galaxies \citep[e.g.,][]{Maller2009}---and not true star formation quenching.
}
Indeed, \citet{Teyssier2012} argued that satellite ejection likely explains the presence of several dwarf galaxies that are beyond the virial radius of the Milky Way and have depleted $\hi$ masses and old stellar populations, based on these galaxies' observed phase-space coordinates.
Finally, \citet{BenitezLlambay2013} and \citet{Bahe2013} examined extended environmental processes on galaxies in SPH simulations; while they found stripping of extended gas (beyond the disk) in the halos of infalling galaxies from ram-pressure in filaments prior to virial crossing, the effect on star formation before virial crossing was negligible for galaxies with $\mstar \gtrsim 10 ^ {9.5} \msun$ (the limit in this paper).

In this paper, we explore observationally and theoretically the evolution of star formation in galaxies near ($1 - 7\,\rthm$) groups and clusters, to understand in detail the physical extent of the environmental dependence of galaxy evolution and the importance of ejected satellites.
We use a galaxy group/cluster catalog constructed from the Sloan Digital Sky Survey (SDSS) together with a cosmological \textit{N}-body simulation, to which we apply the same group-finding algorithm for robust comparison of model results to observations.
This combination allows us to pursue a more rigorous and statistically significant investigation, including accounting for the importance of correlated structure (neighboring halos).
We test the importance of star formation quenching in ejected satellites by employing the same empirically motivated and calibrated model for satellite SFR evolution that we developed in \citet{Wetzel2013}.
To summarize our main result, we find that the quenching of star formation in ejected satellites is consistent with proceeding in the same manner as for satellites that remain within their host halo, and therefore, ejected satellites can account for essentially all of the observed environmental dependence of galaxy quenching beyond the virial radius of massive host halos.

This paper represents the fourth in a series of five.
In \citet{Tinker2011}, hereafter \citetalias{Tinker2011}, we described our SDSS galaxy sample, presented our method for observationally identifying galaxy groups/clusters, and showed that galaxy quiescent fractions are essentially independent of large-scale environment beyond their host halo.
In \citet{Wetzel2012}, hereafter \citetalias{Wetzel2012}, we used our SDSS group catalog to examine in detail the SFR distribution of satellite galaxies and its dependence on stellar mass, host halo mass, and halo-centric distance, finding that the SFR distribution is bimodal in all regimes.
In \citet{Wetzel2013}, hereafter \citetalias{Wetzel2013}, we combined the above results with a cosmological \textit{N}-body simulation to constrain the star formation histories and quenching timescales of satellites, showing that their star formation must evolve in the same manner as central galaxies for $2 - 4 \gyr$ (depending on stellar mass) after falling into a host halo, but that once quenching starts, it occurs rapidly, with a SFR fading e-folding time of $< 1 \gyr$.
Finally, in \citet{Wetzel2013b}, hereafter \citetalias{Wetzel2013b}, we will use the detailed orbital histories from our simulation to test physical mechanisms responsible for satellite quenching.

Before proceeding, we define some nomenclature.
We refer to galaxies with low SFR (see \sect{galaxy_catalog_sdss}) as being `quiescent', and `quenching' is the physical process of SFR fading rapidly from actively star-forming to quiescent, under the ansatz that once a satellite is quenched it remains so indefinitely.
For our galaxy group catalog, `group' means any set of galaxies that occupy a single host halo, regardless of its virial mass, and we use `host halo' as a more general term for group or cluster.
Finally, we cite all masses using $h = 0.7$ for the Hubble parameter.

%===================================================================================================
\section{Methods} \label{sec:method}

Here, we briefly describe our observational galaxy sample and cosmological simulation.
For full details, see Papers I and II for the former and \citetalias{Wetzel2013} for the latter.

%=================================================
\subsection{Observational catalogs from SDSS} \label{sec:sdss}

%========================
\subsubsection{Galaxy catalog from SDSS} \label{sec:galaxy_catalog_sdss}

Our galaxy sample is based on the New York University Value-Added Galaxy Catalog \citep{Blanton2005a} from Data Release 7 of SDSS \citep{Abazajian2009}.
Stellar masses are from the {\tt kcorrect} code of \citet{BlantonRoweis2007}, assuming a \citet{Chabrier2003} initial mass function (IMF).
We construct two volume-limited samples of galaxies with $M_r - 5 \log(h) < -18$ and $-19$, which go out to $z = 0.04$ and 0.06, from which we identify stellar mass complete limits of $5 \times 10 ^ 9$ and $1.3 \times 10 ^ {10} \msun$, respectively.
We use specific star formation rates, $\ssfr = \sfr / \mstar$, based on the current release\footnote{
\tt http://www.mpa-garching.mpg.de/SDSS/DR7/
} 
of the spectral reductions of \citet{Brinchmann2004}, in which $\sfr$ is a galaxy's star formation rate and $\mstar$ its stellar mass.
We define galaxies with $\ssfr < 10^{-11} \yrinv$ as `quiescent', based on the bimodal nature of the SSFR distribution (see \citetalias{Wetzel2012}).

%========================
\subsubsection{Galaxy group catalog from SDSS} \label{sec:group_catalog_sdss}

We identify groups of galaxies that occupy the same host halo using a modified implementation of the group-finding algorithm of \citet{Yang2005a, Yang2007}.
For our group catalog, we define host halos as spherical overdensities containing 200 times the average matter density of the Universe ($\Delta = 200$m).
%For our group catalog, we define host halos such that the average matter density inside of the virial radius, $R_\Delta$, is $\Delta\,\times$ the average matter density of the Universe, $\bar{\rho}_{\rm m}$: $\mthm = 200 \frac{4}{3} \pi \bar{\rho}_{\rm m} \rthm ^ 3$.
As \citetalias{Tinker2011} outlined, we assign host halo virial masses, $\mhalo$, to groups by matching the abundance of halos above a given dark matter mass to the abundance of groups above a given total stellar mass, $n(> \mhalo) = n(> M_{\rm star,\,group})$, using the host halo mass function from \citet{Tinker2008} based on a flat, $\Lambda$CDM cosmology of $\Omega_{\rm m} = 0.27$, $\Omega_{\rm b} = 0.045$, $h = 0.7$, $n_{\rm s} = 0.95$ and $\sigma_8 = 0.82$.
The center of a group is given by its most massive galaxy, which we call the `central' galaxy.
Every group contains one central galaxy and can contain any number (including zero) of less massive `satellite' galaxies.

%=================================================
\subsection{Simulation} \label{sec:simulation}

%========================
\subsubsection{Simulation properties} \label{sec:simulation_property}

To track the evolution of satellites from first infall to final merging/disruption across a broad range of host halo masses, we employ a dissipationless, $N$-body simulation, using the TreePM code of \citet{White2002} with flat, $\Lambda$CDM cosmology of $\Omega_{\rm m} = 0.274$, $\Omega_{\rm b} = 0.0457$, $h = 0.7$, $n = 0.95$ and $\sigma_8 = 0.8$, evolving $2048 ^ 3$ particles of mass $1.98 \times 10^8 \msun$ with a Plummer equivalent smoothing of $3.5 \kpc$ in a $357 \mpc$ box, and storing 45 outputs spaced evenly in $\ln(a)$ from $z = 10$ to 0.
This simulation first was presented in \citet{White2010}.

%========================
\subsubsection{Halo and subhalo tracking} \label{sec:subhalo_tracking}

In the simulation, we identify `host halos' using the Friends-of-Friends (FoF) algorithm \citep{Davis1985} with a linking length of $b = 0.168$ times the average inter-particle spacing, which groups particles bounded by an isodensity contour of $\sim 100 \times$ the average matter density.
Within host halos, we identify `subhalos' as overdensities in phase space through a 6-dimensional FoF algorithm (FoF6D).
For both host halos and subhalos, we keep all objects with at least 50 particles, and we define the center and velocity via the most bound particle.
We assign to each (sub)halo a `child' (sub)halo at the next simulation output, based on its 20 most bound particles.
We define a `central' subhalo as being the most massive subhalo in a newly formed host halo, which almost always corresponds to the subhalo at the minimum of the potential well.
A subhalo retains its central demarcation until it falls into (is linked by the FoF algorithm to) a more massive host halo, becoming a `satellite' subhalo.
Thus, every sufficiently bound halo hosts one central subhalo and can host any number of satellite subhalos.

We assign to each subhalo a maximum mass, $\mmax$, as given by the maximum host halo mass, $\mhalo$, that it ever had as a central subhalo, motivated by the expected correlation of this quantity with galaxy stellar mass (see below).
For a central subhalo, $\mmax$ almost always corresponds to its current halo mass, the primary exception being ejected satellites (see below).
For a satellite, $\mmax$ almost always corresponds to its halo mass shortly before first infall.
Also, as demonstrated in \citet{WetzelWhite2010}, simulations at our resolution scale can resolve and track massive satellite subhalos past the point at which the galactic stellar component that they host would (start to) be stripped, merge with the central galaxy, or otherwise be disrupted, and we use their scheme by removing satellites with $\mbound / \mmax < 0.007$, in which $\mbound$ is the bound subhalo mass as determined by the FoF6D subhalo finder.

Finally, to ensure that the simulation's output time spacing of $\sim 500 \myr$ at $z < 1$ does not bias measurements of satellite orbital properties, such as pericentric distance, we measure the orbital distance and velocity for each satellite in a continuous manner by numerically integrating its orbit between outputs, assuming energy and angular momentum conservation in a spherical NFW \citep{Navarro1997} potential given by the host halo's mass and concentration.

%========================
\subsubsection{Defining satellite infall and ejection} \label{sec:infall_ejection}

We define `first infall' of a subhalo as when it first becomes a satellite in a more massive host halo, if it remains a satellite for at least two consecutive outputs.
The latter criterion avoids cases of temporary bridging if a subhalo briefly passes through the outskirts of a host halo.
After experiencing first infall, if a satellite orbits outside of its host halo, becoming (again) a central subhalo in a distinct halo, we consider it to be an `ejected satellite'.
Thus, an ejected satellite appears to be a central subhalo in an instantaneous sense, though as we will demonstrate, it is more appropriate to continue to consider it a satellite.
Finally, if an ejected satellite grows in halo mass by $> 50\%$ after ejection, we define it to be a newly formed halo and discard it from the ejected population, though as we will show, this criterion affects $< 2\%$ of ejected satellites, regardless of satellite mass.

%========================
\subsubsection{Galaxy group catalog from simulation} \label{sec:catalog_simulation}

Under the ansatz that a galaxy resides at the center of each dark matter subhalo, we assign stellar mass using subhalo abundance matching \citep[SHAM;][]{ValeOstriker2006, Conroy2006}, a method that assumes a one-to-one mapping that preserves rank ordering between subhalo $\mmax$ and galaxy $\mstar$, such that $n(> \mmax) = n(> \mstar)$.
Here, one assigns $\mstar$ to subhalos empirically through an observed stellar mass function (SMF) that is recovered, by design.
SHAM has succeeded in reproducing many observed galaxy statistics, including spatial clustering, satellite fractions, cluster luminosity functions, and luminosity-velocity relations \citep[e.g.,][]{Conroy2006, Berrier2006, Yang2009, WetzelWhite2010, TrujilloGomez2011, Reddick2013}.
We use the SMF from \citet{LiWhite2009}, based on the same SDSS NYU-VAGC sample as our catalog, including the same $K$-correction and IMF.
We apply SHAM to the simulation at $z = 0.05$, implementing a log-normal scatter of 0.15 dex in $\mstar$ at fixed $\mmax$ as suggested by many observations \citep[e.g.,][]{Yang2008, More2009, WetzelWhite2010, Leauthaud2012, Reddick2013, Cacciato2013}.

To make robust comparisons with the SDSS group catalog, we produce a `simulation mock group catalog' by applying the same group-finding algorithm to the simulation.
While we base our models of SFR evolution on true satellite/central demarcation in the simulation, we effectively `observe' our model results through this group catalog, using the same host halo definition and including observational effects, such as redshift-space distortions, that affect group purity/completeness as well as satellite/central assignment.

%===================================================================================================
\section{Star formation in galaxies near groups and clusters} \label{sec:sfr_observe}

We first examine observationally the SFRs of galaxies in and near cluster-mass host halos.
We improve our analysis from \citetalias{Wetzel2012}, in which we examined all galaxies regardless of their redshift at a projected distance from a host halo, by examining only galaxies that lie in close proximity to a given host halo via applying a line-of-sight velocity cut of $\Delta v = \pm 2\,\vthm$, in which $\vthm = \sqrt{G \mthm / \rthm}$ is the virial velocity.
(For reference, $\vthm(10 ^ {14} \msun) = 550 \kms$.)

\begin{figure}
\centering
\includegraphics[width = 0.99 \columnwidth]{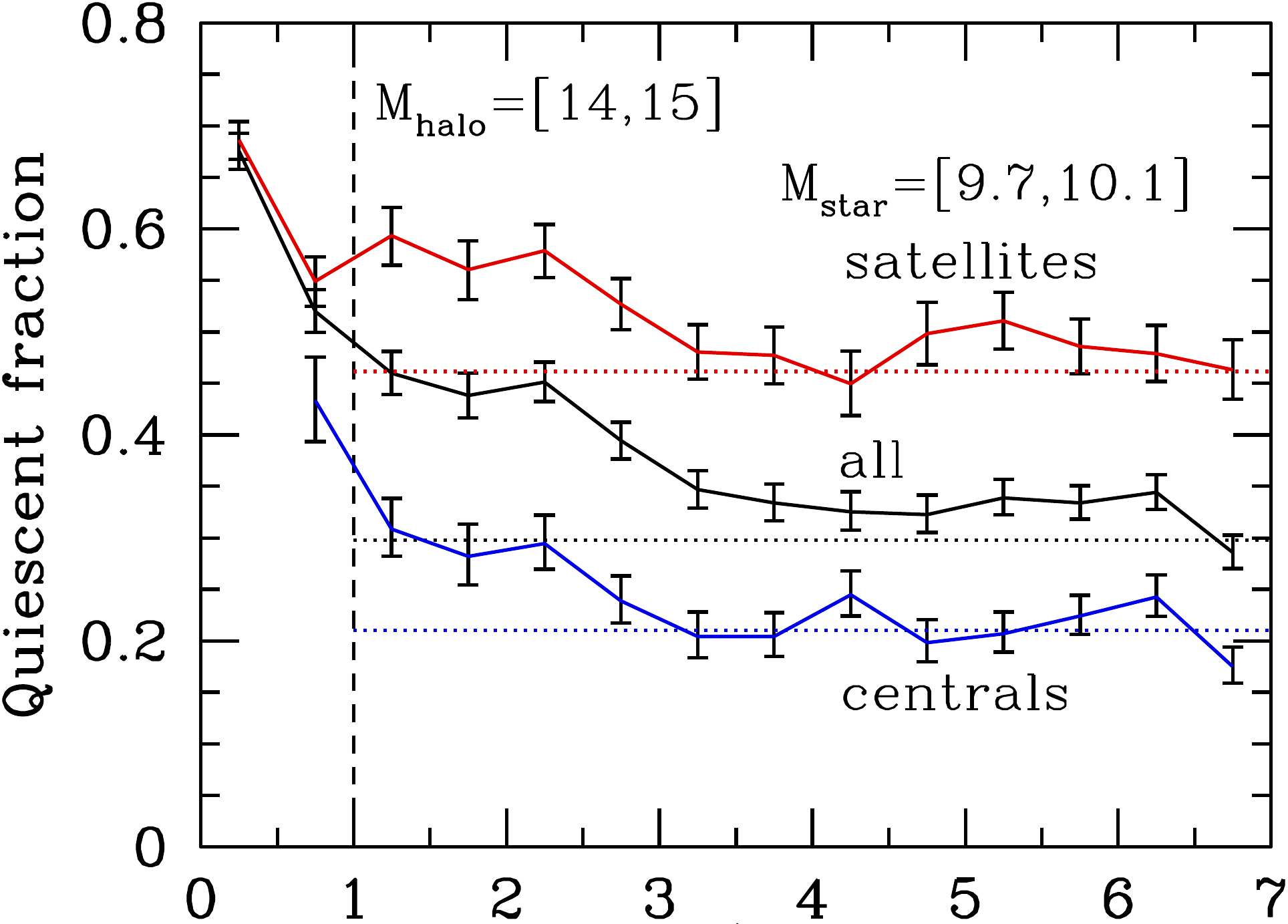}
\includegraphics[width = 0.99 \columnwidth]{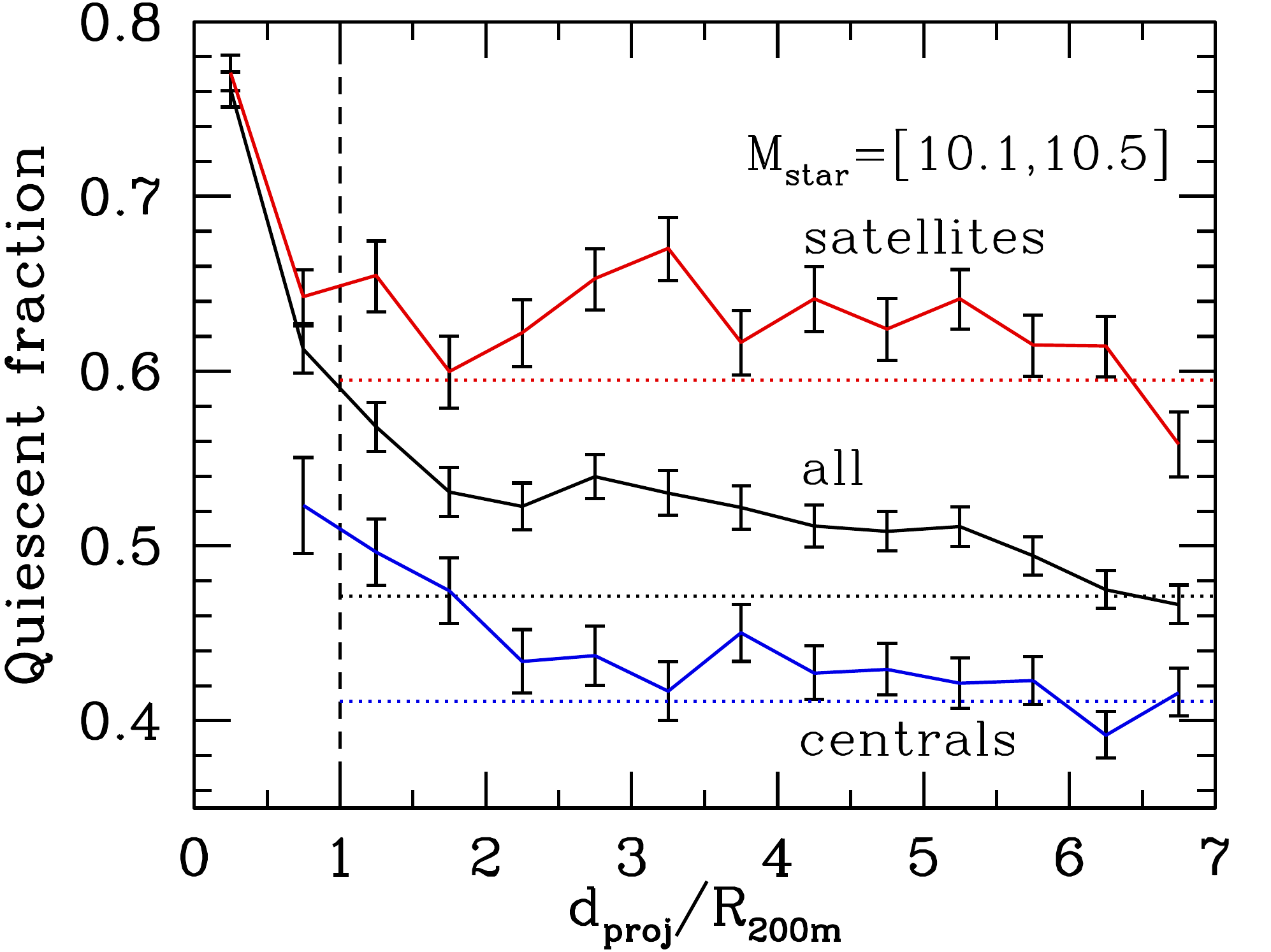}
\caption{
From SDSS: fraction of galaxies that are quiescent ($\ssfr < 10^{-11} \yrinv$) versus projected distance to center of cluster-mass host halos, for galaxies that lie within line-of-sight $\Delta v = \pm 2\,\vthm$ of the host halos.
Panels show bins of galaxy stellar mass, while different curves show galaxies decomposed by type: all (black), satellites (both within the cluster-mass host halos and within neighboring host halos; red), and neighboring central galaxies (blue).
Solid curves show average value in the distance bin, while dotted lines show average across the entire SDSS sample.
Error bars show 68\% confidence interval of the average for a beta distribution.
Considering all galaxies, an enhanced quiescent fraction extends out to $6\,\rthm$, while central galaxies exhibit a strong enhancement out to $\approx 2.5\,\rthm$.
(In all figures, we abbreviate mass labels by removing `$\log$' and `$M_\odot$', such that $\mhalo = \left[14, 15\right]$ means $\log \left(\mhalo / \msun \right) = \left[14, 15\right]$.)
} \label{fig:qu.frac_v_dist.vir_observe}
\end{figure}

\fig{qu.frac_v_dist.vir_observe} shows the fraction of galaxies that are quiescent as a function of projected halo-centric distance, $\dproj$, in two different bins of stellar mass in/around cluster-mass host halos.
Curves show the average quiescent fraction, stacking all host halos in the given range of $\mthm$, and error bars show 68\% confidence interval on the average, as given by a beta distribution \citep{Cameron2011}.
Considering all galaxies, regardless of type (solid black curves), those at $\dproj < \rthm$ have the highest likelihood of being quiescent, but a clear enhancement, beyond the cosmic average of all galaxies in the stellar mass bin (dotted lines), persists out to $\sim 6\,\rthm$ \citep[see also][]{Hansen2009, vonderLinden2010}.

What causes this enhancement out to such large distances, well beyond the turn-around (maximum) distance from which matter accretes onto these clusters?
To address this, we use our group catalog to decompose these galaxies physically by type: satellite (solid red curves) and central (solid blue curves).
(Note that some central galaxies appear within $\rthm$ as viewed in projection.)
This decomposition reveals two trends that cause the enhanced quiescent fraction for all galaxies out to $6\,\rthm$.
First, structure is correlated: massive host halos are more likely to be found near each other.\footnote{
For the cluster-mass host halos in \fig{qu.frac_v_dist.vir_observe}, the median \textit{projected} distance (given our velocity cut) to the center of the nearest host halo with $\mthm > 10 ^ {14, 13, 12} \msun$ is $\dproj / \rthm = 4.7$, 1.4, 0.6, respectively.
In real space, as measured in our simulation, the median $d / \rthm$ values are roughly twice as large.
}
Thus, the likelihood that a galaxy is in another massive host halo decreases with distance from a cluster, so the fraction of galaxies that are satellites, and the average host halo mass of those that are, decreases with distance.
Because the quiescent fraction is higher for satellites than central galaxies, and for satellites it is higher in more massive host halos (\citetalias{Wetzel2012}), correlated structure can cause a gradient in the quiescent fraction with distance even if there were no gradients for central and satellite galaxies separately.

However, \fig{qu.frac_v_dist.vir_observe} also shows another, more interesting trend: central galaxies, which reside in their own host halo, exhibit a strongly enhanced quiescent likelihood out to $\approx 2.5\,\rthm$.
While this enhancement is strongest around cluster-mass host halos, central galaxies near lower mass host halos show similar behavior, and a non-zero enhancement beyond the cosmic average for central galaxies extends out to $\approx 5\,\rthm$ in essentially all cases (see \fig{qu.frac_v_dist.vir_model} in \sect{sfr_evolution}).

What causes this enhanced quiescent fraction in central galaxies near massive host halos?
It potentially could be an artifact of interloping galaxies caused by redshift-space distortions or some other aspect of the group-finding algorithm.
However, as we will show using our simulation mock group catalog in \sect{sfr_evolution}, such observational effects cannot account for this strong excess (in agreement with Y. \citealt{Wang2009b}).
This leaves two feasible physical causes:
\begin{enumerate}
\renewcommand{\labelenumi}{(\alph{enumi})}
\item Strong environmental processes might extend well beyond the (spherical) virial radius of host halos, quenching central galaxies \textit{prior} to their first crossing inside.
\item The observed enhancement could originate from `ordinary' environmental quenching within a host halo's virial radius and propagate to larger distances via `ejected' satellites, which orbit beyond the virial radius after falling inside, becoming (once again) central galaxies in a distinct halo.
\end{enumerate}

Given its simplicity, we postulate that (b) most naturally explains the observed trends, as we will subsequently examine and test.
We will discuss (a) in \sect{robustness}.

%===================================================================================================
\section{Orbital evolution of ejected satellites} \label{sec:orbit}

\begin{figure*}
\centering
\includegraphics[height = 0.26 \textwidth]{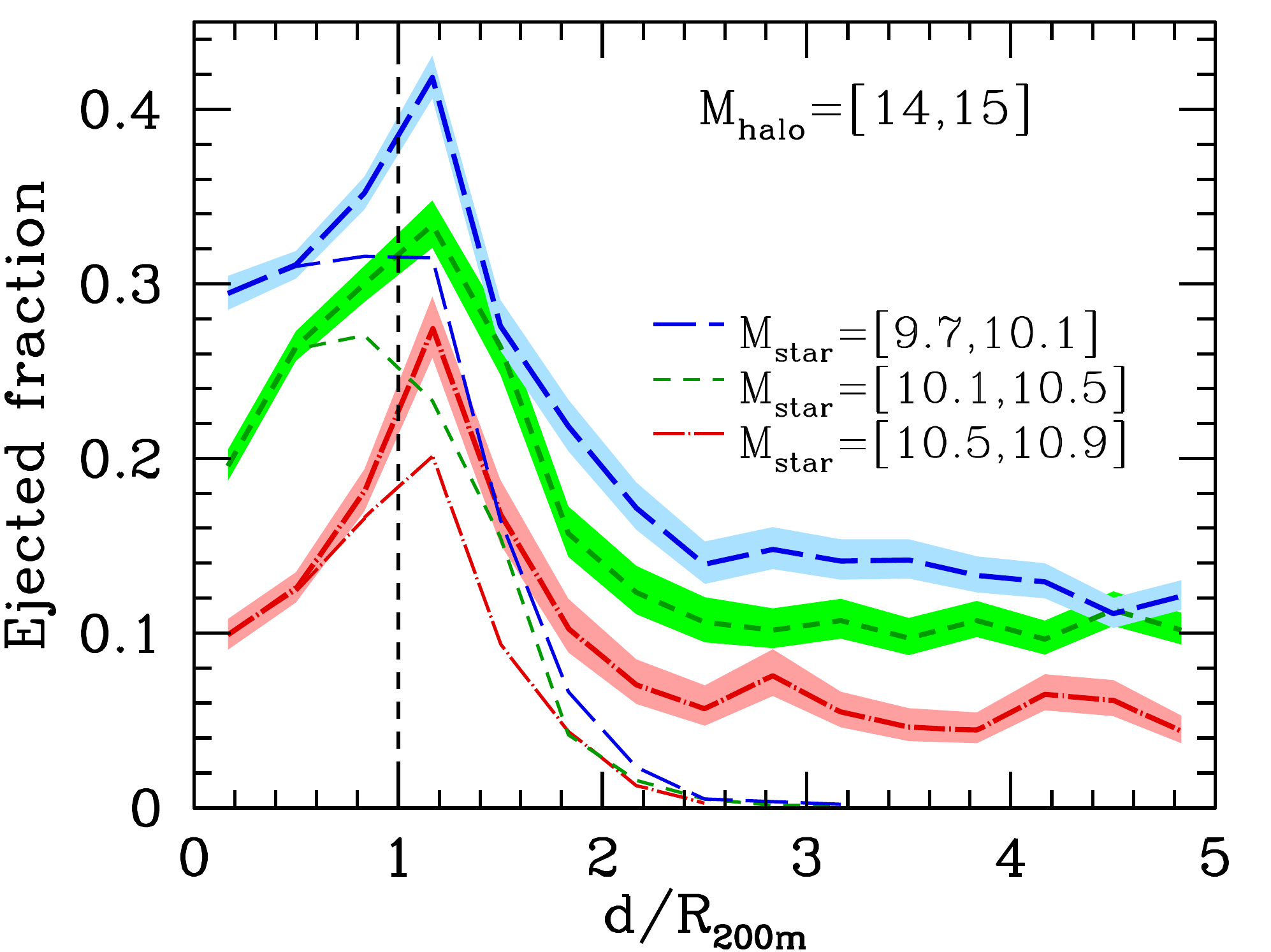}
\includegraphics[height = 0.26 \textwidth]{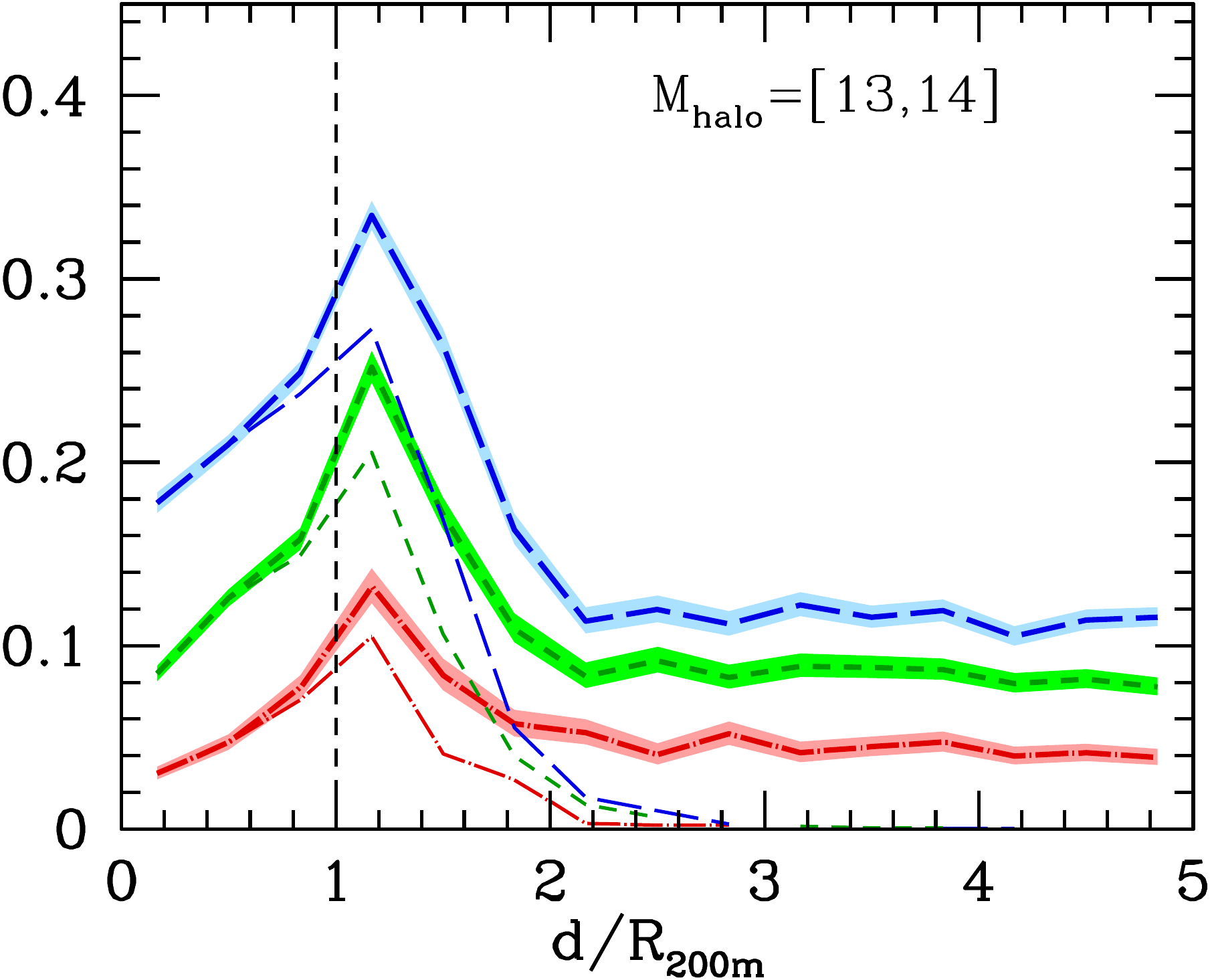}
\includegraphics[height = 0.26 \textwidth]{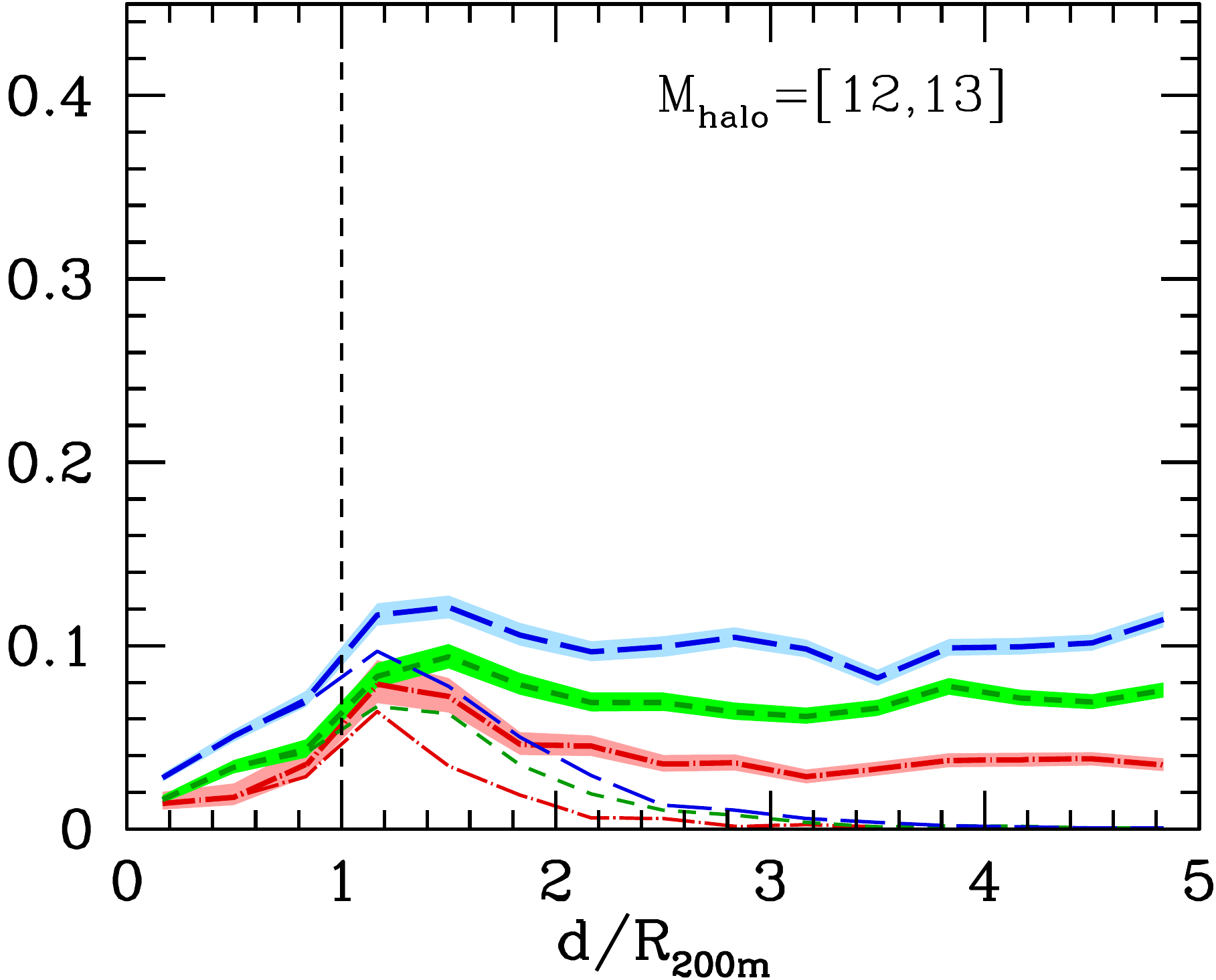}
\caption{
From simulation: fraction of galaxies that have been ejected versus (real-space) halo-centric distance at $z = 0$.
Panels show bins of virial mass of the host halos on which these profiles are stacked, while different curves show bins of galaxy stellar mass.
Values at $d > \rthm$ show the fraction of all central galaxies that are ejected satellites, $\nejected(d) / \ncen(d)$, while values at $d < \rthm$ show the fraction of satellites that experienced an ejected phase and have since fallen back in, $\nejected(d) / \nsat(d)$.
Thin curves (without shading) include just satellites that were ejected from the host halos on which these profiles are centered, while thick curves (with shading) include all ejected satellites, including those ejected from neighboring host halos.
Shaded regions show 68\% confidence interval of the average.
Ejection is more common for less massive galaxies and in/around more massive host halos, being most significant at $1 - 2.5\,\rthm$, though extending to much larger distances because of satellites ejected from neighboring host halos.
} \label{fig:ejected.frac_v_dist.vir}
\end{figure*}

In \citetalias{Wetzel2013}, we showed that a significant fraction of satellites inside a host halo spent time in an ejected phase beyond the virial radius before falling back in.
In this section, we investigate in more detail the orbital histories of ejected satellites, building on previous work but focusing on orbital properties that are (potentially) relevant to quenching star formation.
While we examine galaxies and host halos in the same mass ranges as in our SDSS catalog, in this section we use the halo catalog directly from our simulation (not from the mock group catalog) and measure halo-centric distance in real-space, $d$, in order to highlight physical trends without projection effects or redshift-space distortions.

%=================================================
\subsection{Fraction of galaxies that have been ejected} \label{sec:ejected_fraction}

We first examine the importance and radial extent of ejected satellites around massive host halos.
\fig{ejected.frac_v_dist.vir} shows what fraction of galaxies at a given (real-space) distance around host halos are ejected.
We show the average value from stacking galaxies around host halos of the given virial mass, and shaded regions shows the 68\% confidence interval on this average.
Values at $d > \rthm$ show what fraction of all central galaxies in the distance bin are ejected satellites, $\nejected(d) / \ncen(d)$.
For reference and continuity, \fig{ejected.frac_v_dist.vir} also shows values at $d < \rthm$, which indicate what fraction of all satellites experienced an ejected phase and fell back in, $\nejected(d) / \nsat(d)$.

The solid curves show satellites that were ejected specifically from the host halos on which these profiles are centered.
Across all galaxy and host halo masses, a non-trivial population of ejected satellites extends to $\approx 2.5\,\rthm$ beyond their host halo, in good agreement with previous simulation results \citep{Gill2005, Warnick2008, Ludlow2009, Wang2009a}.
This maximum distance is approximately as expected based on the turn-around distance, $d_{\rm ta}$, of infalling matter, if it conserves energy as it orbits back to apocenter, $d_{\rm apo}$, after infall \citep{Mamon2004}.
We do find that a small fraction ($\sim 10\%$) of ejected satellites extend beyond $1.1\,d_{\rm ta}$, likely caused by an orbital energy boost during a multi-body encounter within the host halo \citep{Sales2007b}.
However, this phenomenon primarily affects satellites with particularly low mass that can be boosted more easily \citep{Ludlow2009}.
The majority (68\%) of satellites in our mass range have $d_{\rm apo} < d_{\rm ta}$, and the vast majority (90\%) have $d_{\rm apo} < 1.1\,d_{\rm ta}$, implying that any such orbital energy boosts are mild for satellites that we examine.

\begin{figure}
\centering
\includegraphics[width = 0.99 \columnwidth]{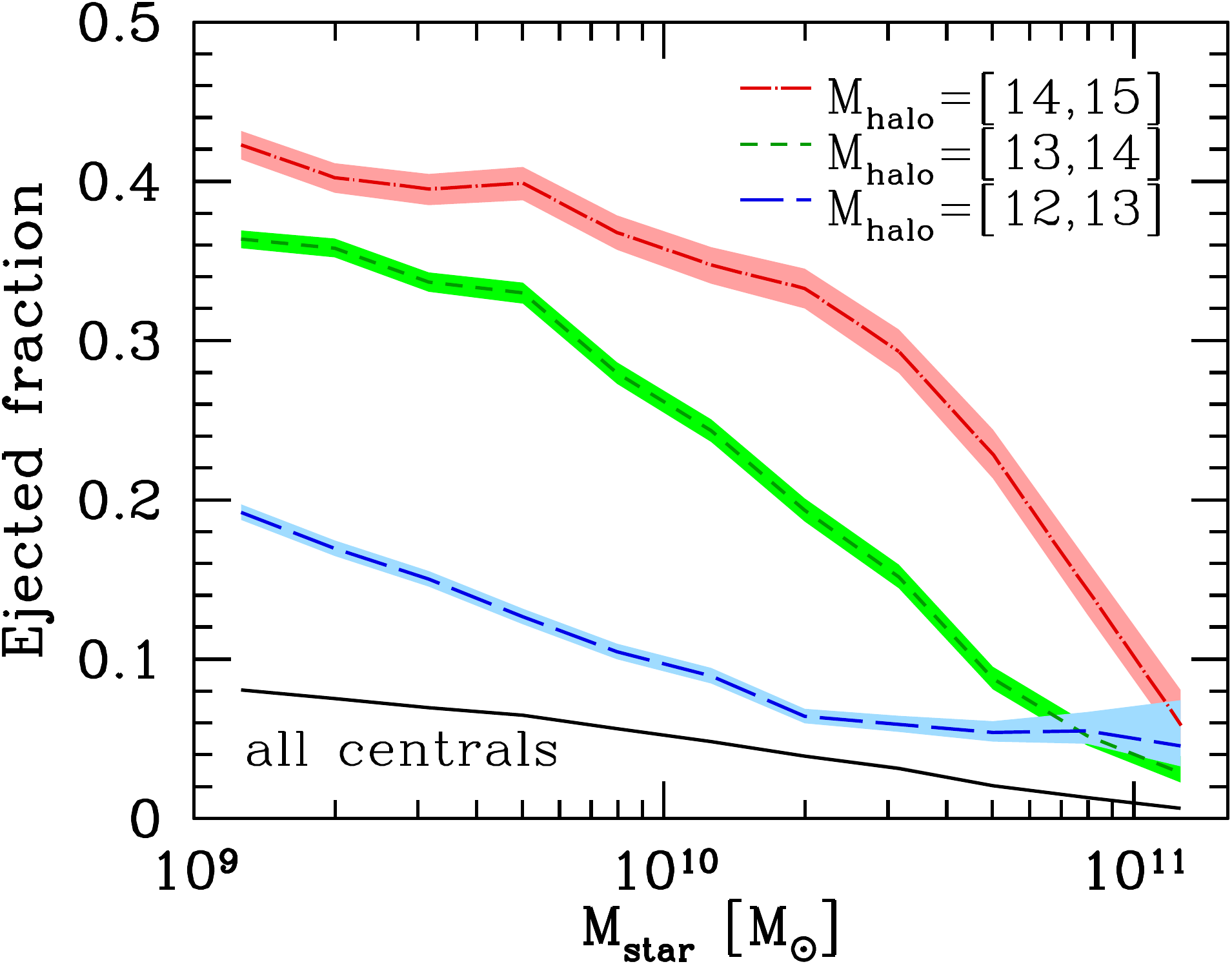}
\caption{
From simulation: fraction of central galaxies that are ejected satellites versus stellar mass at $z = 0$.
Solid curve shows fraction with respect to \textit{all} central galaxies, $\nejected / \ncen$, while dashed curves show fraction within $2.5\,\rthm$ of host halos of the given virial mass, $\nejected(1 - 2.5\,\rthm, \mthm) / \ncen(1 - 2.5\,\rthm, \mthm)$.
Satellite ejection is more common for less massive galaxies and near more massive host halos.
} \label{fig:ejected.frac_v_m.star}
\end{figure}

By contrast, the dashed curves in \fig{ejected.frac_v_dist.vir} show the ejected fraction \textit{regardless} of which host halo the satellites passed through.
Thus, it also includes satellites that were ejected from \textit{neighboring} host halos.
In examining this \textit{total} fraction, a strong excess still persists out to $2.5\,\rthm$, but it declines only gradually before reaching the cosmic average (given in \fig{ejected.frac_v_m.star}) at $\sim 10\,\rthm$.
Thus, while ejected satellites are most common near massive host halos, they are cosmically ubiquitous.
Because separating ejected satellites based on which host halo they passed through observationally is highly non-trivial, if not impossible, one \textit{must} consider this total population when comparing to observations.

\fig{ejected.frac_v_m.star} shows more directly how the fraction of central galaxies that are ejected satellites varies with galaxy stellar mass.
First, the solid curve shows what fraction of \textit{all} central galaxies are ejected satellites.
This fraction increases at lower stellar mass, though it is always $< 10\%$ down to $10 ^ 9 \msun$ \citep[in agreement with][]{Wang2009a}.
Thus, ejected satellites are of modest importance to the overall population of central galaxies.
However, as \fig{ejected.frac_v_dist.vir} showed, ejected satellites are particularly common near massive host halos, and the dashed curves in \fig{ejected.frac_v_m.star} show the ejected fraction for central galaxies that are within $1 - 2.5\,\rthm$ of host halos of the given virial mass.
This ejected fraction strongly increases with host halo mass and decreases with galaxy mass.
Most likely, these mass dependencies arise from the increased efficiency of dynamical friction `braking' of orbital velocity when the ratio of satellite to host halo mass is smaller \citep{BoylanKolchin2008, Jiang2008}.
Note that 40\% of all central galaxies with $\mstar < 10 ^ {10} \msun$ near cluster-mass host halos are ejected satellites.

%=================================================
\subsection{Minimum pericentric distance} \label{sec:distance}

\begin{figure}
\centering
\includegraphics[width = 0.99 \columnwidth]{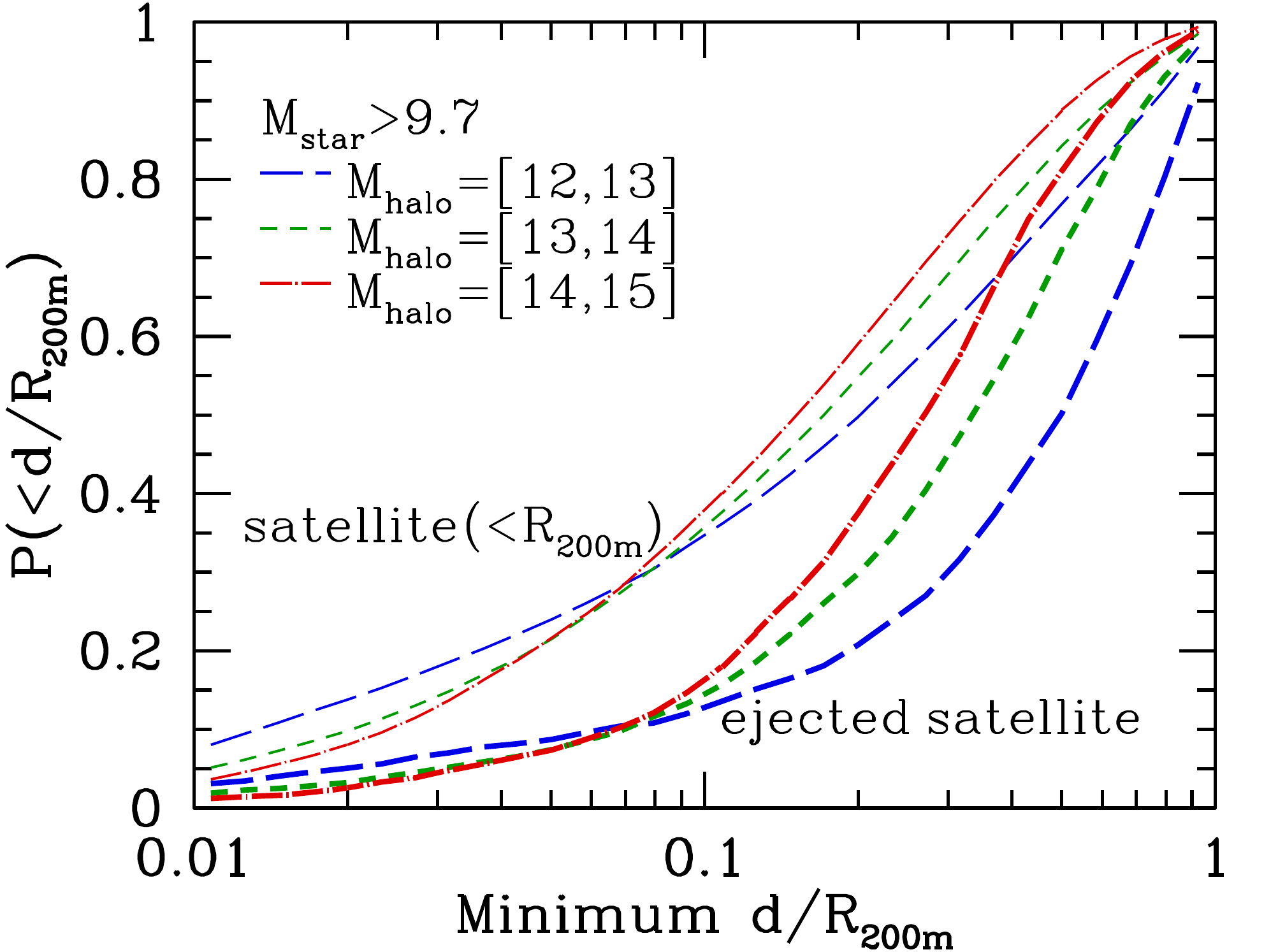}
\caption{
From simulation: cumulative distribution of minimum halo-centric distance to which satellites at $z = 0$ have orbited, in bins of host halo virial mass.
Thick curves show satellites that currently are ejected (beyond $\rthm$), while for reference, thin curves show all satellites that currently are within $\rthm$.
The median pericenter for ejected satellites is $0.25 - 0.5\,\rthm$, being monotonically smaller in more massive host halos.
Ejected satellites almost always experience a single pericentric passage.
} \label{fig:dist.vir.min_distr.cum}
\end{figure}

We next examine how far into their host halo these ejected satellites orbited, in order to understand the feasibility of their star formation being affected. 
We measure a satellite's $d / \rthm$ at each simulation output and identify the minimum ratio experienced since first infall, using orbital integration as given in \sect{subhalo_tracking}.
\fig{dist.vir.min_distr.cum} shows the cumulative distribution of minimum $d / \rthm$ to which different satellite populations at $z = 0$ have orbited.
Solid curves show ejected satellites, which currently are beyond $\rthm$, while for reference, dashed curves show all satellites that currently are within $\rthm$.
Across the entire sample, the median pericentric passage of ejected satellites is $\sim 1 / 3\,\rthm$, sufficiently deep into the halo's potential well to make it feasible that the majority have experienced quenching process(es).
As \fig{dist.vir.min_distr.cum} shows, the median for all satellites within the virial radius is $\sim 0.15\,\rthm$, so by comparison, ejected satellites have not orbited especially deeply.
Though, it may be unfair to compare ejected satellites to all satellites, given that almost half of all satellites \textit{currently} lie within $1 / 3\,\rthm$.
If instead we compare the pericenter distribution of ejected satellites to those currently at $0.9 - 1\,\rthm$, they are almost identical (not shown).
Thus, the majority of ejected satellites are not on orbits with particularly unusual pericenters, they simply have sufficient orbital energy to bring them back beyond the virial radius.

\fig{dist.vir.min_distr.cum} also shows that satellites in more massive host halos typically orbited to somewhat smaller $d / \rthm$, with the median value for ejected satellites varying from $0.25 - 0.5\,\rthm$ across our host halo mass range.
We find no dependence of any of these distributions on satellite mass.
Finally, the pericentric distances experienced by ejected satellites in \fig{dist.vir.min_distr.cum} are almost always ($\sim 90\%$) the result of a \textit{single} pericentric passage between first infall and ejection, also noted previously \citep{Gill2005}.
The $\sim 10 \%$ of ejected satellites that experienced multiple pericentric passages typically are on highly radial orbits and experienced several passages in/out of the host halo.

%=================================================
\subsection{Orbital timescales} \label{sec:time}

Our primary goal is to test whether ejected satellites experience the same quenching of star formation as `normal' satellites that remain within a host halo.
As \citetalias{Wetzel2013} showed, SFR in normal satellites evolves unaffected for $2 - 4 \gyr$ after first infall, depending on stellar mass, after which it is quenched rapidly, with an e-folding time of $< 0.8 \gyr$.
Thus, we now examine the orbital timescales of ejected satellites in the context of these quenching times.

\begin{figure}
\centering
\includegraphics[width = 0.99 \columnwidth]{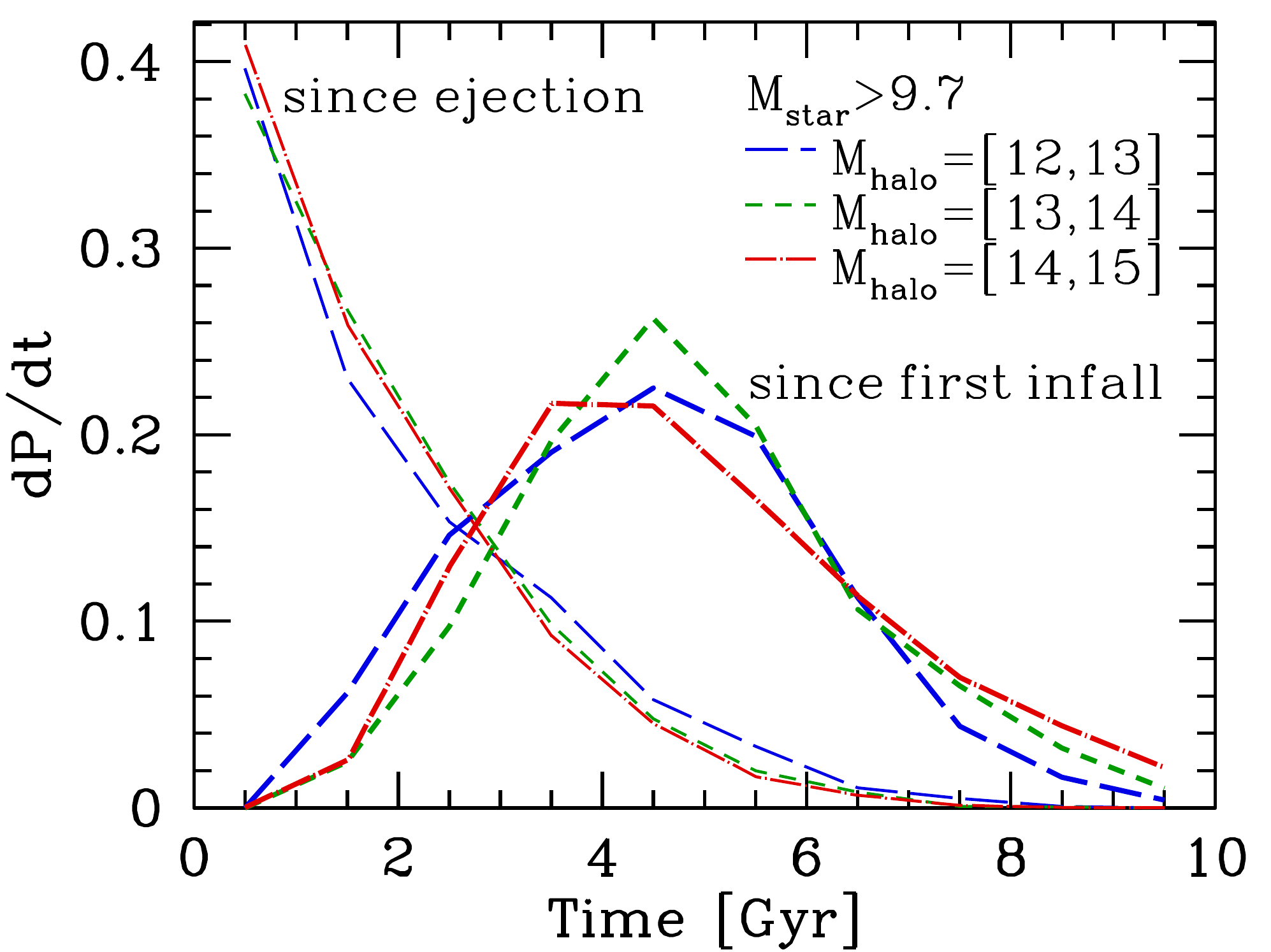}
\caption{
From simulation: distribution of time since first infall (thin curves) and time since ejection (thick curves) for ejected satellites at $z = 0$, in bins of the virial mass of the host halo that they passed through.
The median time since first infall is $4.8 \gyr$ and since ejection is $1.4 \gyr$.
These orbital times do not depend significantly on either host halo mass or satellite mass and are comparable to satellite quenching timescales.
} \label{fig:times_distr}
\end{figure}

\fig{times_distr} shows the distribution of time since first infall (dashed) and time since ejection (solid) for ejected satellites at $z = 0$.
We find little dependence of these orbital times on the virial mass of the host halo that they passed through, as shown, and no dependence on galaxy mass (not shown).
Across all ejected satellites, the median time since ejection is $1.4 \gyr$ and since first infall is $4.8 \gyr$.
The distribution of time since ejection falls off rapidly, meaning that relatively few satellites are on (nearly) unbound orbits that keep them beyond the virial radius for a long time.
Instead, almost all ejected satellites remain bound to their host halo.
For ejected satellites that are crossing back inside the virial radius at $z = 0$, we find that the median time spent in an ejected phase is $3.6 \gyr$.

We also find that the median amount of time that ejected satellites spent within their host halo is $2.9 \gyr$ across our sample, again with no significant dependence on galaxy or host halo mass.
This time-as-satellite is similar to the virial crossing time ($2\,\rthm / \vthm$) at the typical redshift when these satellites fell in ($z \sim 0.5$), as expected given that the vast majority experienced a single pericentric passage, and in good agreement with \citet{Wang2009a}.\footnote{
While \citet{Wang2009a} found a shorter timescale for more massive satellites, they conjectured that this arose from a higher liklihood of transient fly-by's for more massive satellites.
Our more conservative selection criterion, which requires an ejected satellite to have remained within its host halo for at least two consecutive simulation outputs, leads to no such mass dependence.
}

Thus, compared to the quenching delay timescales of satellites from \citetalias{Wetzel2013}, the typical time-as-satellite is similar, and the typical time since first infall is considerably longer.
As such, most ejected satellites easily have the potential to be quenched in the same way as normal satellites.

%=================================================
\subsection{Evolution of halo mass} \label{sec:m.halo_evolution}

\begin{figure}
\centering
\includegraphics[width = 0.99 \columnwidth]{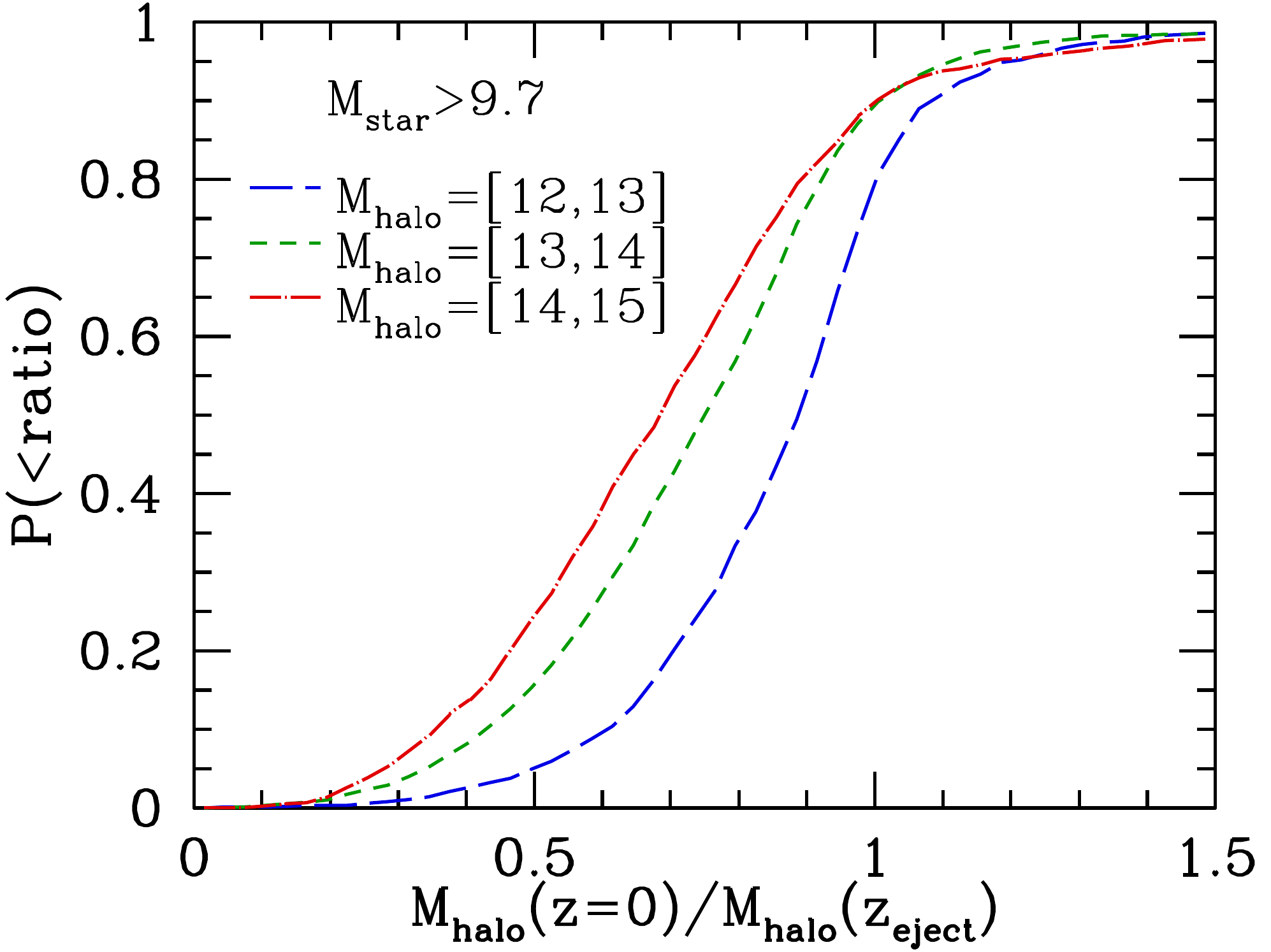}
\caption{
From simulation: cumulative distribution of the fraction of halo mass that an ejected satellite has at $z = 0$ relative to what it had immediately after ejection, in bins of the virial mass of the host halo that it passed through.
Ejected satellites experienced somewhat stronger stripping in/near more massive host halos, with median fractions ranging from $0.7 - 0.9$.
Only $10 - 20\%$ of ejected satellites have gained mass since ejection; most continue to experience halo mass stripping, with stronger stripping near more massive host halos.
} \label{fig:m.halo.frac_distr.cum}
\end{figure}

Having shown that most ejected satellites orbited relatively deeply into their host halo and have been satellites for a sufficiently long time to be (at least potentially) quenched, we next explore the evolution of their (sub)halo mass after ejection.
If an ejected satellite's (sub)halo starts to accrete mass again after ejection, in the same way as non-ejected halos, this accretion feasibly could lead to renewed star formation, even if quenching already occurred.
Alternately, if ejected satellites typically do not accrete mass, then they are likely to remain quenched in the same way as normal satellites.

To demonstrate the evolution of halo mass after ejection, \fig{m.halo.frac_distr.cum} shows the cumulative distribution of the fraction of halo mass that an ejected satellite has at $z = 0$ relative to what it had immediately after ejection.
Different curves show satellites ejected from host halos of different virial mass.
Across all host halos, only $\sim 10\%$ of ejected satellites gained halo mass since ejection.
Instead, ejected satellites almost always continue to lose mass, both from tidal stripping in the dynamically hot environment surrounding a massive host halo \citep[e.g.,][]{Hahn2009} and from the delayed response of particles from more impulsive unbinding closer to pericenter.
\fig{m.halo.frac_distr.cum} shows that stronger stripping occurs near more massive host halos, with the typical mass loss since ejection being 40\% near clusters \citep[in good agreement with][]{Gill2005, Warnick2008}, but only 10\% near $\mthm = 10 ^ {12 - 13} \msun$.\footnote{
While halo mass stripping is significant after ejection, the total amount of stripping since first infall is dominated by the stripping that occurs as a satellite subhalo near pericenter.
Across our sample of ejected satellites, the loss after ejection accounts for only $\sim 17\%$ of the total (sub)halo mass loss since first infall.
}
We find no significant dependence on satellite mass.
This continued halo mass loss strongly suggest that the vast majority of ejected satellites do not restart star formation from renewed gas accretion after ejection, but rather continue to be affected and quenched after ejection in the same way as satellites within a host halo.
These results also have important implications for the stellar-to-halo mass relation of ejected satellites, as we will explore in \sect{m.star_m.halo_relation}.

%=================================================
\subsection{Summary} \label{sec:orbit_summary}

To summarize this section, the orbits of ejected satellites extend to $\approx 2.5\,\rthm$ beyond their host halo, though because of neighboring host halos and correlated structure, an enhanced total population extends out to $\sim 10\,\rthm$.
Lower mass satellites and those associated with more massive host halos are more likely to experience an ejected phase. 
After first infall, these satellites almost always experience a single pericentric passage before ejection, typically to $\sim 1 / 3\,\rthm$, which is similar to satellites that are just within $\rthm$.
Thus, the typical time that ejected satellites spent within the host halo, $\sim 3 \gyr$, is approximately the virial crossing time, and almost all remain on bound orbits and re-enter the host halo after $\sim 3.5 \gyr$.
For satellites that currently are ejected, their typical time since first becoming a satellite is $\sim 5 \gyr$, longer than the $2 - 4 \gyr$ quenching delay time of satellites, and most (90\%) continue to lose halo mass after ejection.
These results strongly suggest that ejected satellites quench and remain quenched in the same way as satellites inside a host halo, which we next test directly.

%===================================================================================================
\section{Quenching of star formation in ejected satellites} \label{sec:sfr_evolution}

In \sect{sfr_observe}, we postulated that ejected satellites could explain the observed enhancement in the quiescent fraction for central galaxies near massive host halos.
We now test this scenario directly by examining two contrasting models for the evolution of SFR in ejected satellites:
\begin{enumerate}
\renewcommand{\labelenumi}{(\alph{enumi})}
\item `satellite-like' evolution, in which the SFRs of ejected satellites are quenched in the same way as normal satellites,
\item `central-like' evolution, in which the SFRs of ejected satellites evolve in the same way as normal (non-ejected) central galaxies.
\end{enumerate}

To test `satellite-like' evolution, we apply the same physically motivated, two-stage model for the evolution of SFR in satellites from \citetalias{Wetzel2013}.
To summarize, we start by obtaining statistically accurate initial SFRs for satellites at their time of first infall via an empirically based parametrization for the evolution of central galaxy SFRs out to $z = 1$, combining our SDSS group catalog with data at higher $z$ from the Cosmic Evolution Survey (COSMOS) \citep{Drory2009} and the All-wavelength Extended Groth strip International Survey (AEGIS) \citep{Noeske2007}.
If a satellite was actively star-forming at its time of first infall, $\tinf$, obtained via the simulation, we parametrize its subsequent evolution as follows.
Its SFR fades gradually, in the same manner as central galaxies, across a `delay' time, $\tqdelay$.
After this, it starts to be quenched, and its SFR fades exponentially across an e-folding time, $\tauqfade$:
\begin{equation} \label{eq:sfr_evolution_sat}
\begin{split}
\sfrsat&(t) = \\
&\begin{cases}
\sfrcen(t) & t < \tqstart \\
\sfrcen(\tqstart) \exp {\left\{ -\frac{t - \tqstart}{\tauqfade} \right\}} & t > \tqstart
\end{cases}
\end{split}
\end{equation}
in which $\tqstart = \tinf + \tqdelay$, and
\begin{equation} \label{eq:sfr_evolution_cen}
\sfrcen(t) = \sfr_0 (t - t_{\rm f}) \exp \left\{ -\frac{t - t_{\rm f}}{\taucen} \right\} \\
\end{equation}
with $t_f$ being the time of initial formation, which we take to be $t(z = 3)$.
We obtain $\sfr_0$ and $\taucen$ in narrow bins of stellar mass using the measured $\mstar$ and $\sfr$ of actively star forming central galaxies in the SDSS group catalog (assuming 40\% stellar mass loss through supernovae and stellar winds by $z = 0$).
As \citetalias{Wetzel2013} showed, this empirically constrained model provides accurate SFR distributions and quiescent fractions for satellites at $z = 0$ across both stellar and host halo mass, with $\tqdelay$ and $\tauqfade$ that depend on satellite mass but not significantly on host halo mass.
(As \citetalias{Wetzel2013b} will show, this model also reproduces accurately the dependence of SFR versus halo-centric distance within host halos.)
We use the same quenching timescales as \citetalias{Wetzel2013}: $\tqdelay = 3.4 \gyr$ and $\tauqfade = 0.8, 0.6 \gyr$ for $\log \left( \mstar / \msun \right) = 9.7 - 10.1, 10.1 - 10.5$, respectively.\footnote{
These timescales, and all results in this paper, are based on our `fiducial' parametrization in \citetalias{Wetzel2013} for the initial quiescent fraction for satellites at $\tinf$.
Using our `alternate' parametrization, along with the associated quenching timescales, does not change any results in this paper significantly.
}

For a contrasting null model, we consider `central-like' evolution, in which ejected satellites have the same instantaneous distribution of SFRs as central galaxies.
Physically, this scenario implies either that SFR in ejected satellites is not affected environmentally, so it evolves in the same way as in normal central galaxies, or that SFR in ejected satellites is quenched environmentally, but it resumes to being the same as in normal central galaxies after ejection.
To implement this model, we proceed as above, but we re-assign SFRs to ejected satellites empirically by drawing randomly from `isolated' central galaxies of the same stellar mass in the SDSS group catalog that do not lie within $3\,\rthm$ in projection of a more massive host halo, to minimize any possible contamination from ejected satellites.

The two models above provide SFRs at $z = 0$ for all satellites in the simulation.
We assign SFRs to central galaxies by drawing randomly from isolated central galaxies of the same stellar mass in the SDSS group catalog.
Thus, we assign SFR to galaxies in the simulation according to their physical satellite versus central demarcation, but we then measure the results using the mock group catalog to compare robustly with SDSS.

\begin{figure}
\centering
\includegraphics[width = 0.99 \columnwidth]{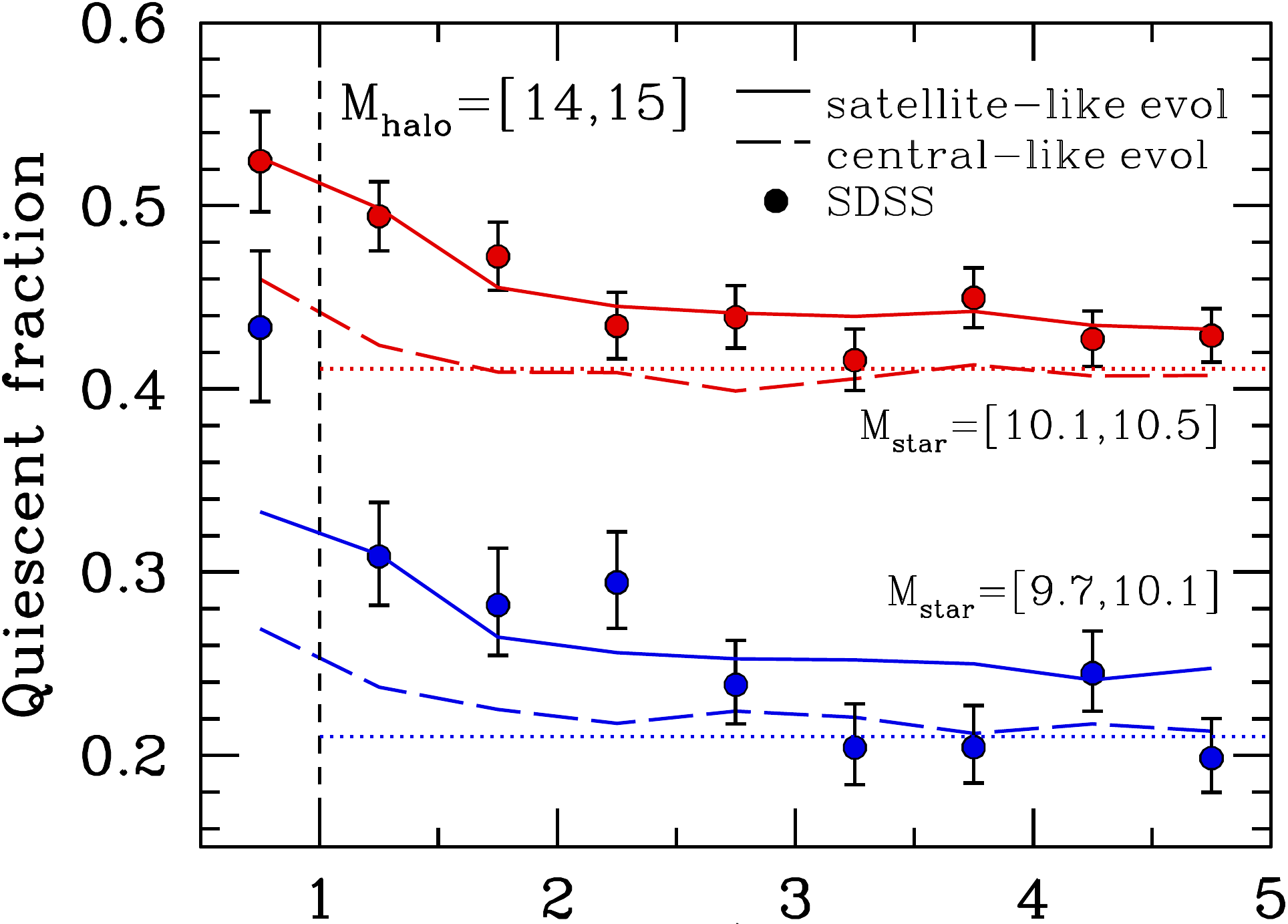}
\includegraphics[width = 0.99 \columnwidth]{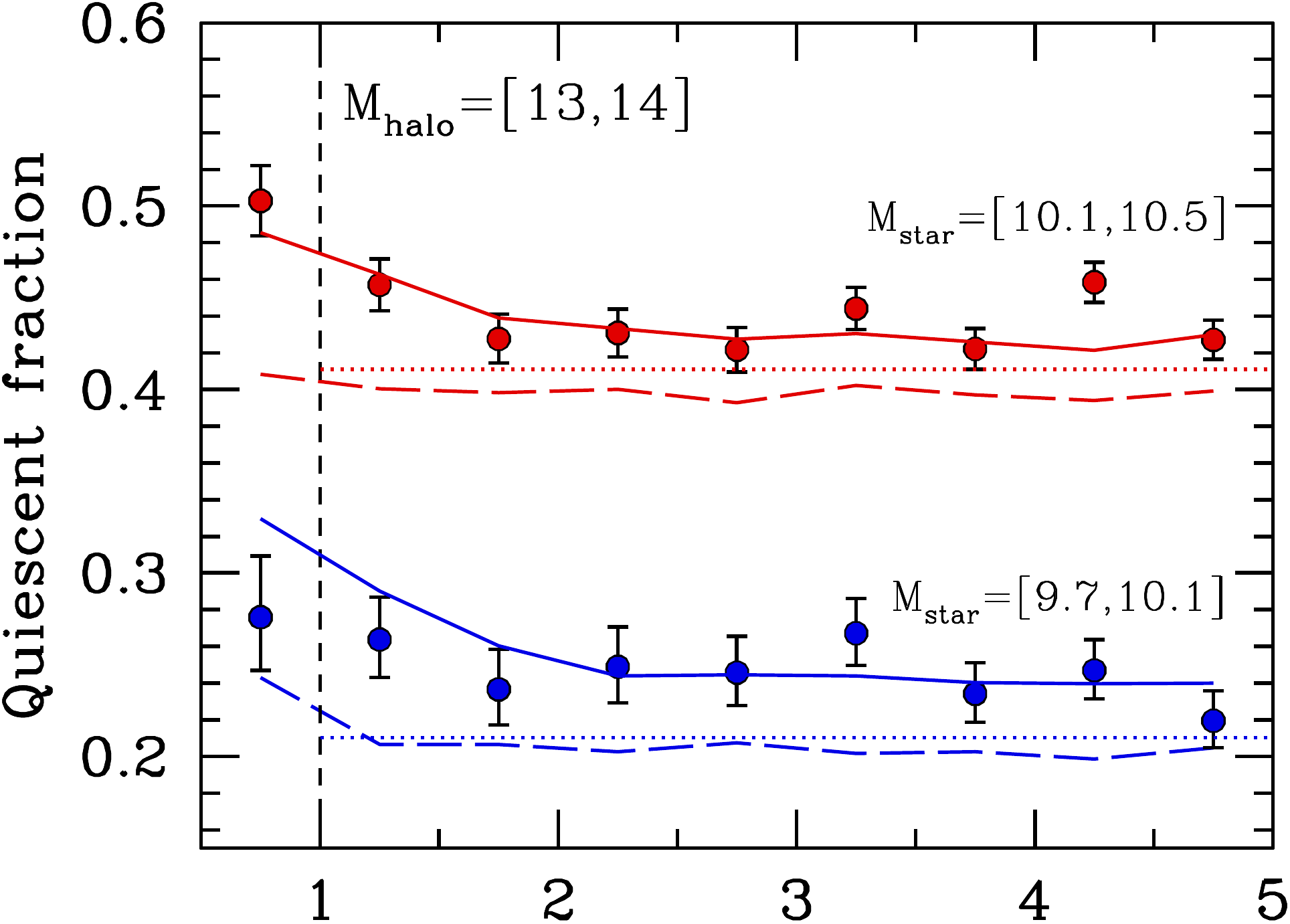}
\includegraphics[width = 0.99 \columnwidth]{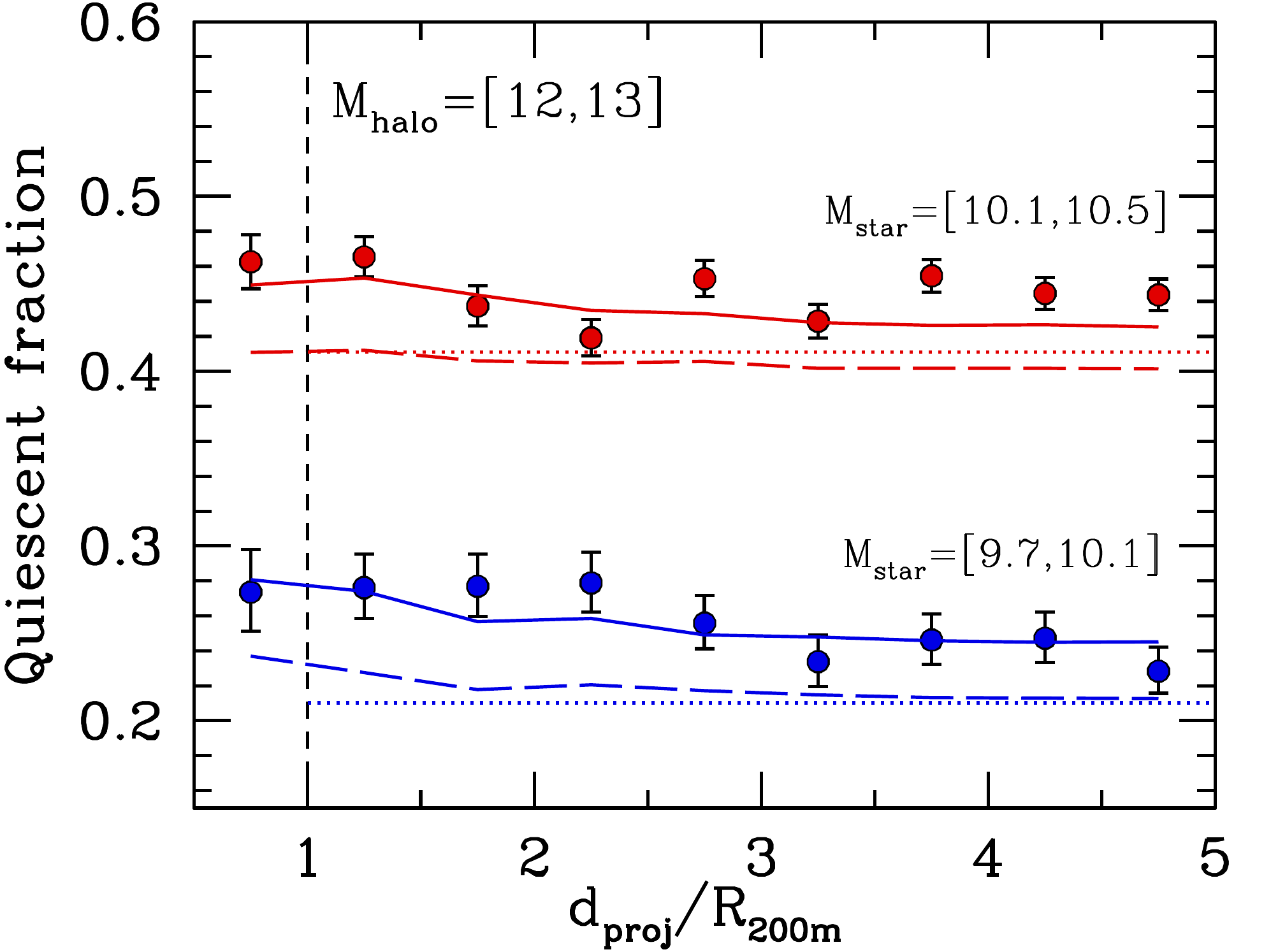}
\caption{
Comparing models to SDSS: fraction of central galaxies that are quiescent versus projected distance to massive host halos.
Panels show bins of host halo mass, different colors show bins of galaxy stellar mass.
From the SDSS group catalog, points show average value in the distance bin, and dotted lines show average for all central galaxies in the catalog (as in \fig{qu.frac_v_dist.vir_observe}).
Curves show models as `observed' through the simulation mock group catalog: if ejected satellites have the same SFR as isolated central galaxies (dashed), or if SFR in ejected satellites evolves and is quenched in the same way as other satellites, according to \eq{sfr_evolution_sat} (solid).
Overall, the latter model successfully can explain the enhanced quiescent fraction in central galaxies out to $5\,\rthm$.
} \label{fig:qu.frac_v_dist.vir_model}
\end{figure}

\fig{qu.frac_v_dist.vir_model} shows the quiescent fraction for central-identified galaxies as a function of projected distance from host halos of the given mass in each panel, comparing the results of the two models above against the SDSS group catalog.
As in \fig{qu.frac_v_dist.vir_observe}, points show the average observed value in the distance bin from SDSS, and dotted lines show the average for all central galaxies of the given stellar mass in SDSS.
Note that, as compared with \fig{qu.frac_v_dist.vir_observe}, \fig{qu.frac_v_dist.vir_model} extends to host halos of much lower mass, where the upturn near $\rthm$ is weaker, but the excess beyond the cosmic average remains strong out to $5\,\rthm$.
Each panel shows two bins of galaxy stellar mass; we find that galaxies of higher mass show less enhancement, as expected from \fig{ejected.frac_v_dist.vir}.

Curves show the two model results as measured through the simulation mock group catalog.
Dashed curves show `central-type' evolution for ejected satellites, in which \textit{all} central galaxies have the same SFR distribution, regardless of formation history.
We note two important points.
First, even when viewed through the mock group catalog, this null model recovers well the average quiescent fraction for central galaxies (dotted lines).
Some upturn occurs at $\rthm$, caused by redshift-space distortions, but it is weak compared to the observed enhancement.
Thus, we conclude that observational effects related to the group finder alone cannot account for the strong enhancement in the quiescent fraction for central galaxies: it is a real physical effect.
Second, because this model recovers the cosmic average, it underestimates the observed quiescent fraction even out to $\approx 5\,\rthm$.
Thus, the importance of quenching in ejected satellites is not confined just to within $2.5\,\rthm$ of host halos; it is critical to understanding the \textit{entire} population of central galaxies.

By contrast, solid curves show `satellite-type' evolution for ejected satellites, in which their SFR is quenched in the same way as normal satellites.
This model provides good overall agreement for the upturn within $2.5\,\rthm$.
Just as importantly, it also correctly enhances the quiescent fractions at larger $\dproj$, because of satellites ejected from nearby host halos, as \fig{ejected.frac_v_dist.vir} showed.

We emphasize that this model for `satellite-type' SFR evolution is simply an extension of our empirically constrained model from \citetalias{Wetzel2013}---which provided excellent agreement for satellites that remain within host halos---to the ejected satellite population.
Furthermore, our hypothesis that ejected satellites are quenched (and remain quenched) is supported physically by their orbital histories, as \sect{orbit} showed.
Thus, given the good agreement of this model with SDSS in \fig{ejected.frac_v_dist.vir}, we arrive at the following main conclusions:
\begin{enumerate}
\renewcommand{\labelenumi}{$\bullet$}
\item SFR evolution and quenching in ejected satellites is consistent with proceeding in the same manner as in normal satellites that remain within their host halo.
\item This quenching of ejected satellites naturally can explain the enhanced quiescent fraction for central galaxies out to $5\,\rthm$ beyond host halos.
\item The success of this model leaves little-to-no room for environmental processes that quench central galaxies beyond a host halo's virial radius, that is, essentially all environmental quenching is consistent with originating within a host halo's virial radius.
\end{enumerate}

%===================================================================================================
\section{$\mstar - \mhalo$ relation for ejected satellites} \label{sec:m.star_m.halo_relation}

\begin{figure}
\centering
\includegraphics[width = 0.99 \columnwidth]{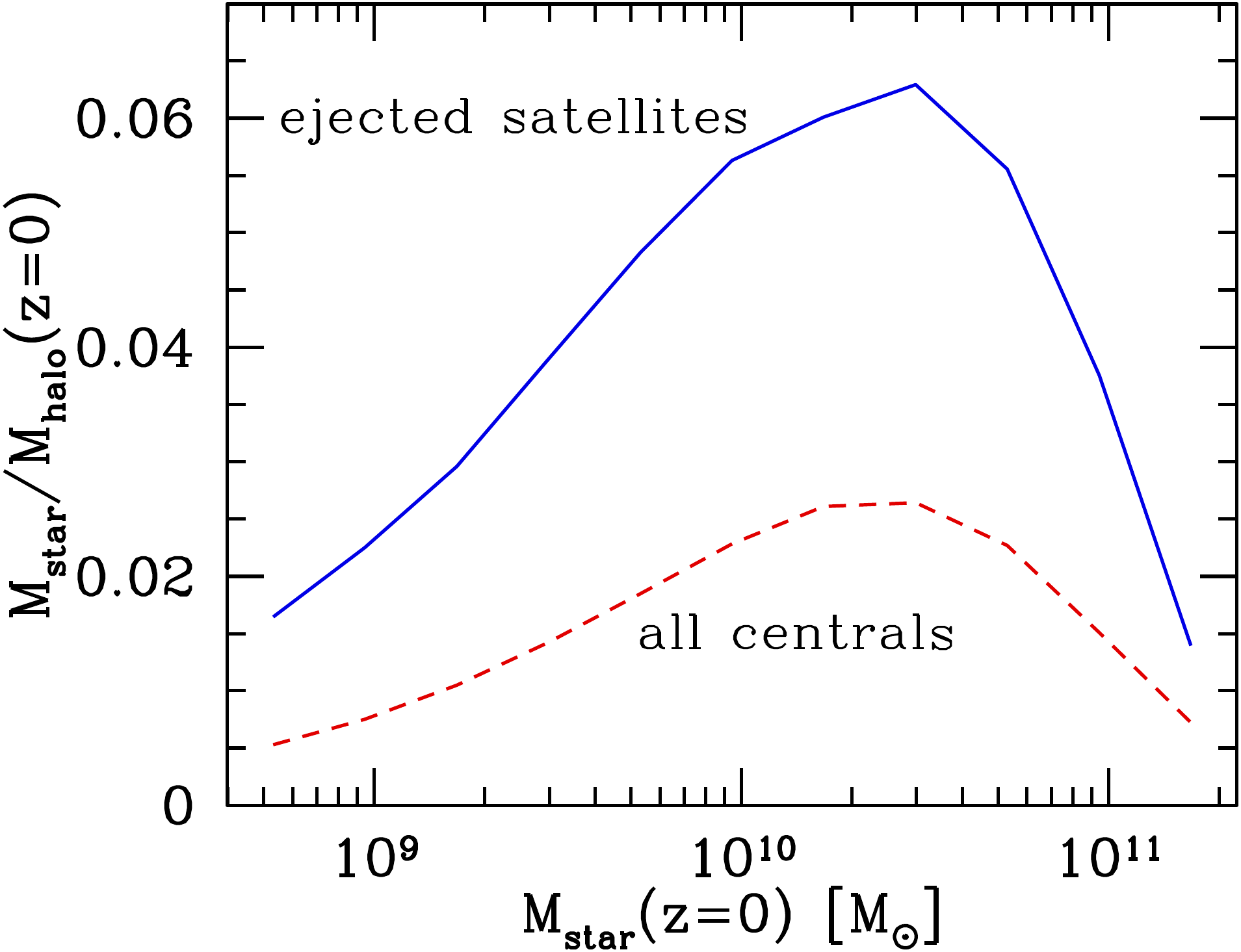}
\caption{
From simulation: median ratio of a central galaxy's stellar mass, $\mstar$, to its (instantaneous) halo virial mass, $\mhalo$, as a function of stellar mass at $z = 0$, for all central galaxies (dashed) and just ejected satellites (solid), as obtained via abundance matching using $\mmax$.
Because of strong stripping of (sub)halo mass after infall and ejection, ejected satellites have $2.5 \times$ higher ratio of $\mstar$ to $\mhalo$ than for all central galaxies.
} \label{fig:m.star_m.halo_rat_v_m.star}
\end{figure}

In addition to star-formation quenching, in this last section we explore the implications of the mass evolution of ejected satellites.
As \sect{m.halo_evolution} showed, because of the tidal stripping of (sub)halo mass that occurs after first infall and continues after ejection, ejected satellites have significantly reduced halo masses as compared with normal (non-ejected) central galaxies of the same stellar mass.
However, because a galaxy is smaller and more tightly bound than its (sub)halo, ejected satellites likely do not experience significant stellar mass loss from tidal stripping during their (typically single) passage through the host halo \citep{Knebe2011}.
Thus, central galaxies that are ejected satellites would have a higher ratio of stellar to halo mass than normal central galaxies.

To illustrate the significance of this effect for the entire population at $z = 0$, \fig{m.star_m.halo_rat_v_m.star} shows median $\mstar / \mhalo$ for all central galaxies in the simulation (dashed) and for just central galaxies that are ejected satellites (solid), as obtained through subhalo abundance matching using $\mmax$ (\sect{catalog_simulation}).
Because of mass stripping, the halo mass of an ejected satellites is typically only $\sim 40\%$ that of the whole central galaxy population of the same stellar mass.
In other words, \textit{ejected satellites systematically have a $\sim 2.5 \times$ higher ratio of $\mstar$ to $\mhalo$ than all central galaxies}, with a somewhat stronger fractional enhancement at lower mass.

\begin{figure}
\centering
\includegraphics[width = 0.99 \columnwidth]{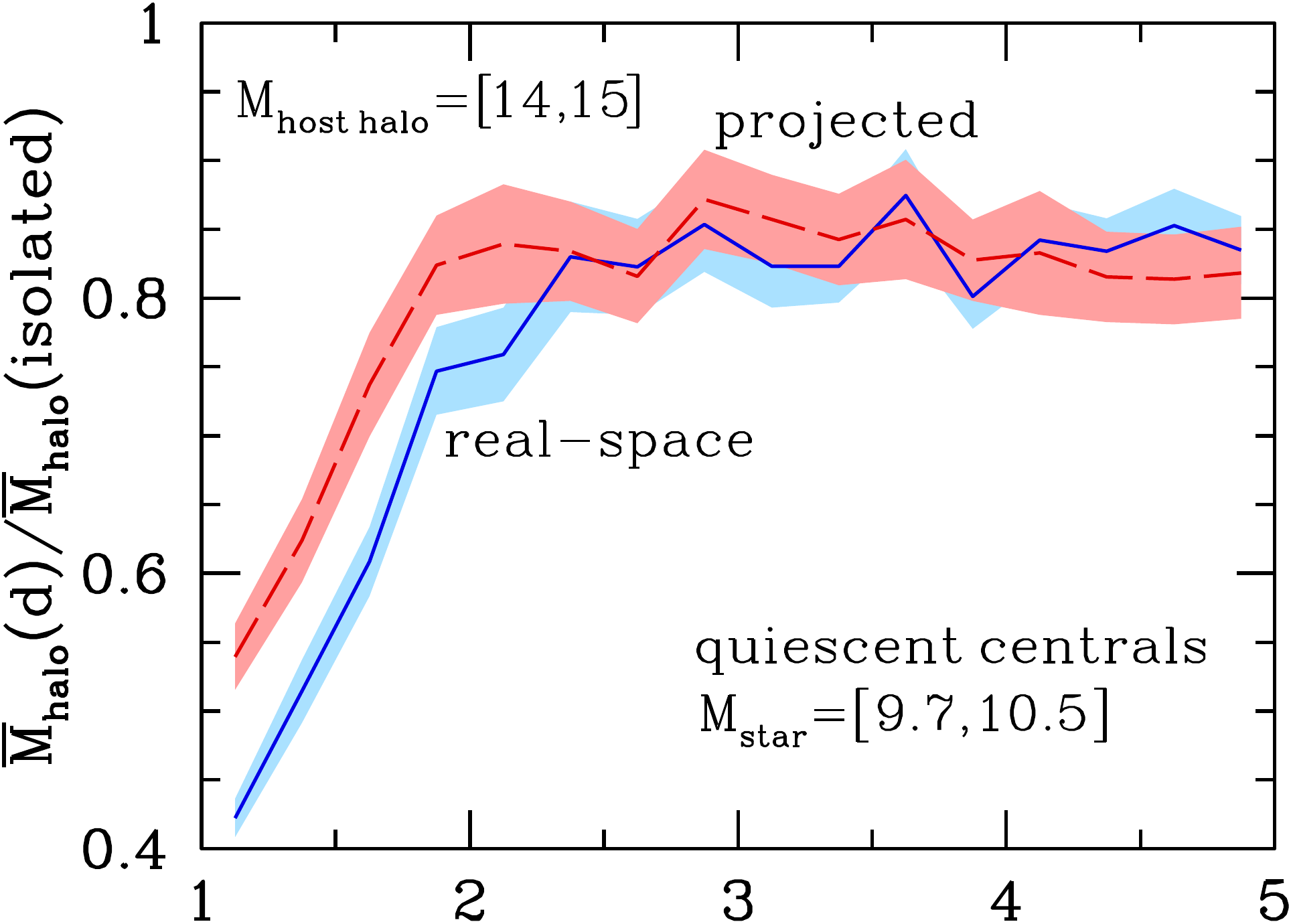}
\includegraphics[width = 0.99 \columnwidth]{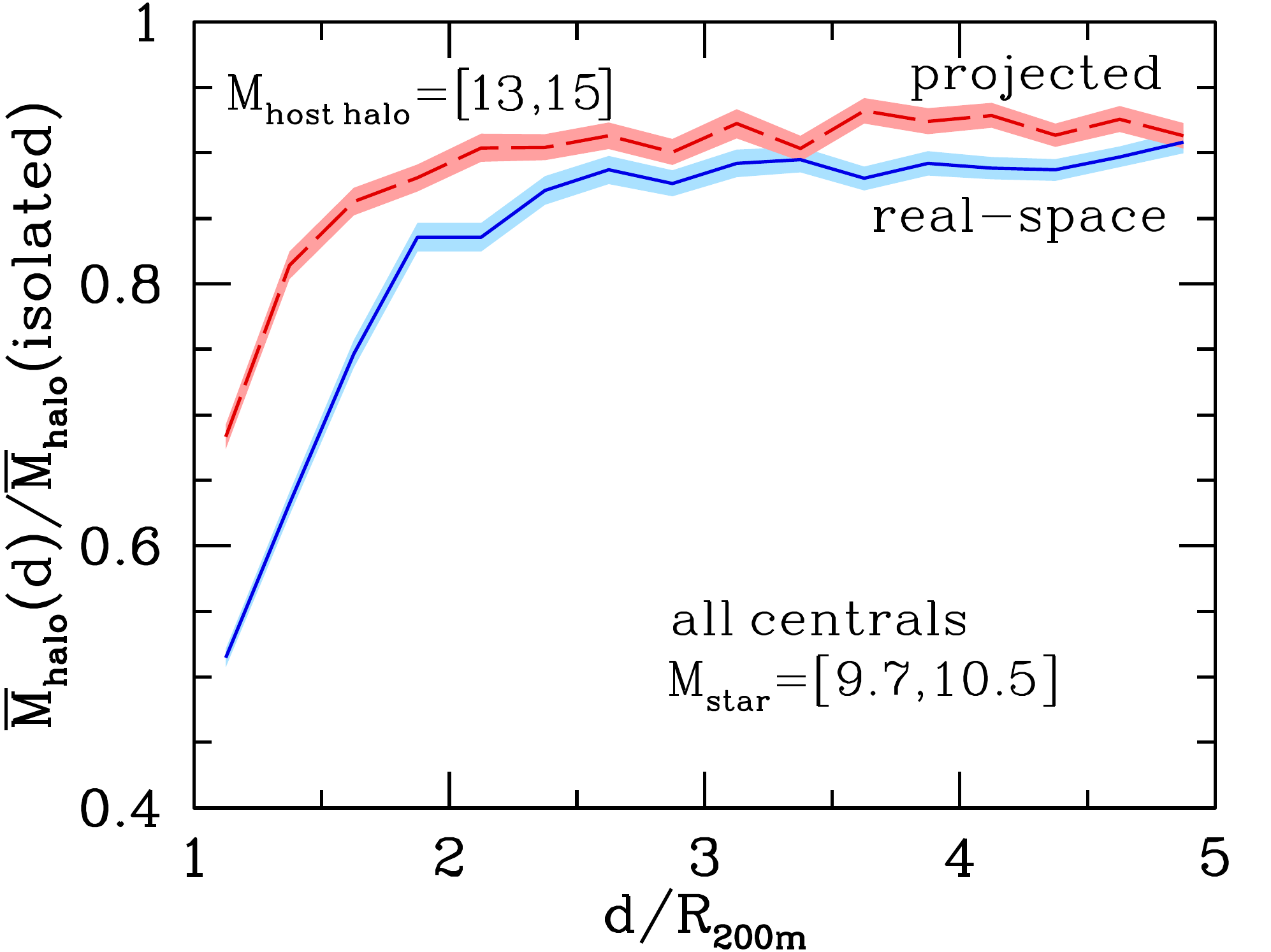}
\caption{
From simulation: average halo mass for central galaxies in the given stellar mass bin versus halo-centric distance at $z = 0$, normalized to the average for isolated central galaxies in the stellar mass bin.
Solid curves show values in real-space from the simulation directly, while dashed curves show values in projection measured using the simulation mock group catalog.
Shaded regions show 68\% confidence interval of the average.
\textbf{Top}: Quiescent central galaxies near cluster-mass host halos, where the signal is strongest.
\textbf{Bottom}: All central galaxies near group- and cluster-mass host halos, where statistics are stronger.
A significant reduction in average halo mass, driven primarily by tidal stripping of ejected satellites, persists out to $\approx 2.5\,\rthm$.
} \label{fig:m.halo_v_dist.vir}
\end{figure}

While \fig{m.star_m.halo_rat_v_m.star} shows the offset in halo mass if one isolates the population of ejected satellites, in reality they are spatially intermixed with normal (first-infalling) central galaxies.
Because of the intrinsic difficulty in separating these populations, observationally measuring this effect requires stacking and averaging both populations, leading to a weaker measured offset.
The population of ejected satellites is most significant near the most massive host halos (\fig{ejected.frac_v_dist.vir}), where halo mass stripping also is strongest (\fig{m.halo.frac_distr.cum}).
Additionally, ejected satellites are more likely to be quiescent than non-ejected central galaxies of the same stellar mass (\fig{qu.frac_v_dist.vir_model}).
Thus, quiescent, low-mass central galaxies near the most massive host halos should provide the strongest measurable reduction in average halo mass.

To examine the potential observability of this effect, we use our `satellite-like' evolution model from \sect{sfr_evolution} to assign SFRs to all galaxies in our simulation.
We then stack all quiescent central galaxies near cluster-mass host halos and measure their average halo mass as a function of halo-centric distance.
\fig{m.halo_v_dist.vir} (top) shows this average halo mass, normalized to the average halo mass of all isolated (do not lie within $3\,\rthm$ of a more massive host halo) central galaxies of the same stellar mass.
We select galaxies with $\mstar = 10 ^ {9.7 - 10.5} \msun$ to match our SDSS sample, but we note that this effect is even stronger at lower stellar mass because of a higher likelihood of being an ejected satellite.

Solid blue curves show the true, underlying average as measured in real space, using the halo catalog directly from the simulation.
At $d \approx \rthm$, where ejected satellites are most common, the average halo mass of all central galaxies is 45\% that of isolated central galaxies.\footnote{
Some tidal stripping does occur as far out as 1.5 virial radii for halos that are on first infall \citep{Hahn2009}.
We find that first-infall centrals within $1.5\,\rthm$ have average halo masses that are $\sim 10\%$ lower than isolated centrals, so this does contribute to \fig{m.halo_v_dist.vir} but is not the dominant effect.
We also checked that the distribution of stellar masses within our chosen range does not vary with distance to create additional signal.
}
This reduction remains particularly strong out to $\approx 2.5\,\rthm$, as expected given the spatial extent of ejected satellites from \fig{ejected.frac_v_dist.vir}.
Similar to \fig{qu.frac_v_dist.vir_model}, the average ratio does not asymptote until $\approx 10\,\rthm$ because of satellites ejected from neighboring host halos, and because ejected satellites are cosmically ubiquitous, the ratio asymptotes to $\approx 0.93$ and not 1.

Dashed red curves show the average as measured in projection using the simulation mock group catalog, imposing a line-of-sight velocity cut of $\pm 300$ km/s (more conservative than before to maximize signal).
While measuring in projection reduces the signal somewhat, it remains strong if using a sufficiently conservative velocity cut.\footnote{
The choice of velocity cut represents a trade-off between purity (intrinsic signal) and completeness (statistics), and the choice should be optimized for each observational measurement.
}

These results suggest that the reduced halo masses for central galaxies near groups/clusters as caused by ejected satellites could be measured observationally via galaxy-galaxy weak gravitational lensing, a technique that can measure accurate halo masses for samples of stacked galaxies \citep[e.g.,][]{Mandelbaum2006}.
However, in applying such a technique to the sample in \fig{m.halo_v_dist.vir} (top), note that stacking all clusters with $\mthm = 10 ^ {14 - 15} \msun$ in our SDSS group catalog leads to $\sim 500$ quiescent galaxies at $\mstar = 10 ^ {9.7 - 10.5} \msun$ in each bin of $\Delta \dproj / \rthm = 1$, a number that is unlikely to yield a discriminating stacked lensing signal from current or near-term surveys \citep{Li2013}.
However, if we consider instead \textit{all} central galaxies near host halos with $\mthm = 10 ^ {13 - 15} \msun$, the number in each such distance bin increases to $\sim 5000$.
\fig{m.halo_v_dist.vir} (bottom) shows average halo masses for this much larger sample.
While the average is 10 - 15 percentage points higher, the order-of-magnitude increase in galaxy count makes such a lensing measurement potentially feasible with existing SDSS data, and upcoming measurements from deeper imaging surveys such as the Dark Energy Survey (DES), Hyper Suprime-Cam (HSC), and the Large Synoptic Survey Telescope (LSST) should provide strong constraints on these halo masses of ejected satellites \citep[e.g.,][]{Li2013}.

Indeed, galaxy-galaxy lensing measurements already show reduced subhalo masses for satellite galaxies \textit{within} individual clusters \citep{Limousin2007, Natarajan2009}, and \citet{Gillis2013} measured reduced (sub)halo masses for galaxies in high-density environments (defined by local galaxy density) from the Canada-France-Hawaii Telescope Lensing Survey (CFHTLenS).
In addition to measurements of subhalo masses \textit{within} groups/cluster, such as these, we emphasize the utility of measuring galaxies \textit{near} groups/clusters: while the (sub)halo masses of ejected satellites are not as tidally truncated as those of satellites at the core of their group/cluster, galaxies near groups/clusters can provide a cleaner lensing signal because their contrast with the background density field is much stronger.
Overall, these results suggest promise in measuring (sub)halo stripping in ejected satellites via galaxy-galaxy lensing, and we will pursue further investigation in future work.

In summary, ejected satellites have systematically biased halo masses, typically only 40\% as high as all central galaxies of the same stellar mass, leading to a $\approx 2.5 \times$ higher $\mstar / \mhalo$ ratio.
This truncation strongly manifests itself in the average halo masses of central galaxies out to $\approx 2.5\,\rthm$ beyond groups/clusters and should be measurable via galaxy-galaxy lensing, particularly with measurements from surveys such as DES, HSC, and LSST.

%===================================================================================================
\section{Summary and Discussion} \label{sec:conclusion}

%=================================================
\subsection{Summary of Results} \label{sec:summary}

Using a galaxy group/cluster catalog from SDSS to decompose galaxies into `centrals' and `satellites', we examined the quiescent fraction for galaxies with $\mstar = 10 ^ {9.7 - 10.5} \msun$ that reside near groups/clusters with $\mthm = 10 ^ {12 - 15} \msun$. 
We postulated that the observed enhancement of the quiescent fraction in central galaxies beyond the virial radius of host halos is caused by ejected satellites, that is, galaxies that fell into a host halo and then orbited back out beyond the virial radius, becoming again central galaxies in their own halo.
We used a cosmological \textit{N}-body simulation to examine their orbital histories and mass evolution, and we examined scenarios for their SFR evolution, which we tested by applying the same group-finding algorithm to the simulation to make robust comparisons with SDSS.
Our primary results are as follows.
\\

\textit{Central galaxies near more massive host halos are more likely to be quiescent.}
Considering all galaxies regardless of type, an enhanced quiescent fraction (beyond the cosmic average at a given stellar mass) extends out to $\sim 6\,\rthm$ around massive host halos.
However, this enhancement is caused partially by satellites that reside in neighboring host halos and the fact that host halos are spatially clustered.
Considering only central galaxies, which reside in their own host halo, an enhanced quiescent fraction extends out to $\approx 5\,\rthm$ and is especially strong within $2.5\,\rthm$.

\textit{A large fraction of central galaxies near massive host halos are ejected satellites.}
The orbits of ejected satellites extend to $\approx 2.5\,\rthm$ beyond their host halo, though because of satellites ejected from neighboring host halos, ejected satellites represent a significant fraction of all central galaxies out to much larger distances.
Satellites of lower mass and those in/near more massive host halos are more likely to experience an ejected phase, a likely result of the decreased efficiency of dynamical friction braking in those regimes.
In the extreme case, ejected satellites represent 40\% of all central galaxies of $\mstar < 10 ^ {10} \msun$ within $2.5\,\rthm$ of clusters.

\textit{The orbital histories of ejected satellites are consistent with satellite-like environmental quenching.}
After first infall, ejected satellites almost always ($\sim 90\%$) experienced a single pericentric passage, typically to $\sim 1 / 3\,\rthm$, spending roughly a virial crossing time, $\sim 3 \gyr$, within their host halo.
Almost all remain on bound orbits and (will) re-enter the host halo $\sim 3.5 \gyr$ after ejection.
Their typical time since first infall is $\sim 5 \gyr$, longer than the satellite quenching timescale, and almost all (90\%) continue to lose (sub)halo mass after ejection and do not re-accrete mass.

\textit{The quenching of star formation in ejected satellites is consistent with occuring in the same manner as in normal satellites, and this can explain the enhanced quiescent fraction for central galaxies near massive host halos.}
We applied the same model for satellite quenching, based on time since first infall, that we developed for normal satellites in \citetalias{Wetzel2013} to ejected satellites.
This simple extension of our model naturally and accurately accounts for the enhanced quiescent fraction in central galaxies out to $\approx 5\,\rthm$.

\textit{Effectively all environmental quenching is consistent with occurring within a host halo's virial radius.}
The success of our simple and natural extension of satellite quenching timescales to ejected satellites leaves little-to-no room for additional environmental effects for quenching central galaxies prior to first virial infall.

\textit{Ejected satellites have significantly biased halo masses, and this is potentially observable.}
The (sub)halos of ejected satellites lose significant mass from tidal stripping after both first infall and ejection.
To the extent that the stellar mass remains unstripped, ejected satellite have a $\approx 2.5$ times higher $\mstar / \mhalo$ ratio (or 60\% lower halo mass) than all central galaxies of the same stellar mass.
A reduction of average halo mass extends to central galaxies $\approx 2.5\,\rthm$ beyond massive host halos and is potentially observable via galaxy-galaxy weak lensing.

%=================================================
\subsection{Discussion} \label{sec:discussion}

%========================
\subsubsection{Comparison with previous work} \label{sec:comparison}

Our results broadly agree with those of previous work on the SFRs of galaxies near groups/clusters \citep{Wang2009b, Mahajan2011, Teyssier2012}, though our approach represents a significant advancement: using a larger galaxy sample and examining a broader range of galaxy and host halo masses, accounting for correlated structure and neighboring host halos through our SDSS group catalog, tracking individual satellite orbital histories in simulation, applying an empirically constrained model for SFR evolution and quenching in satellites, and robustly comparing our model results with SDSS through a simulation mock group catalog.
For instance, our results, based on a much larger sample of more massive galaxies, firmly support the scenario as proposed by \citet{Teyssier2012} that satellite ejection explains the presence of dwarf galaxies that are beyond the virial radius of the Milky Way and have depleted $\hi$ masses and old stellar populations.

Our results also connect with previous work on mass stripping in ejected satellites.
\citet{Knebe2011} examined the mass-to-light ratios of ejected (backsplash) versus infalling central galaxies in an SPH simulation of a Local Group analogue.
They also found that ejected satellites have a higher ratio of stellar to halo mass than infalling central galaxies, but interestingly, they measured the mass-to-light ratio \textit{within} the extent of the galactic stellar component (farthest star particle).
Their result suggests that, in addition to stripping halo mass, tidal forces might heat the orbits of stars in the discs of ejected satellites, which could provide another means to differentiate observationally ejected satellites from infalling centrals based on stellar concentration and/or kinematics.

%========================
\subsubsection{Robustness of results} \label{sec:robustness}

We have argued that ejected satellites, which passed within the virial radius of a more massive host halo, can account for essentially all galaxy quenching beyond the virial radius.
However, we cannot \textit{entirely} rule out environmental processes that extend beyond the virial radius, quenching central galaxies prior to crossing inside.
Our model parametrizes the quenching of star formation in ejected satellites in terms of time since first infall (first virial radius crossing), with an initial `delay' time over which star formation is unaffected, followed by rapid quenching.
From \citetalias{Wetzel2013}, this model successfully accounts for the SFR distribution of satellite galaxies in groups/clusters as a function of both galaxy and host halo mass, and, as \citetalias{Wetzel2013b} will show, successfully describes the dependence of the quiescent fraction on halo-centric distance \textit{within} groups/clusters.
However, it may be possible that other models, in which the quenching process starts at a farther distance, work as well.
We will test and compare other physical models for satellite quenching in \citetalias{Wetzel2013b}.

None-the-less, our model, in which the environmental quenching process starts at first virial infall, can explain the SFRs of central galaxies beyond groups/clusters, and we argue that this is the simplest and most natural explanation for the extended environmental quenching, particularly given what we already know about satellite quenching within the virial radius.
The results of \citet{BenitezLlambay2013} and \citet{Bahe2013}, both based on SPH simulations, also broadly support this picture for SFR in galaxies in our mass range.
While they found some stripping of extended gas (beyond the disk) in the halos of infalling galaxies prior to virial crossing from ram-pressure in filaments, the effect on star formation before virial crossing was negligible for galaxies in our mass range, in agreement with our results.
Thus, to the extent that the extension of our model to ejected satellites is correct, the agreement of our model with SDSS means that any environmental effects beyond the virial radius are sub-dominant for galaxies in our mass range.
Though, \citet{BenitezLlambay2013} and \citet{Bahe2013} did see stronger reduction of star formation prior to virial crossing for galaxies of significantly lower mass, which could be tested in the future through observations of lower mass galaxies.

We did consider two tests to isolated potential environmental effects on galaxies within our mass range before first infall.
First, we examined the relative velocities (with respect to the host halo) of ejected versus first-infalling halos at $d / \rthm = 1 - 2.5$, finding that the median velocity components of ejected satellites can be lower than that of first-infalling halos ($\sim 30\%$ for tangential, $\sim 10\%$ for radial), especially near massive clusters.\footnote{
We thank the reviewer for suggesting this test.
}
However, the 68\% scatter in velocity of both populations is large (factor of 3 - 4, see \citealt{Wetzel2011}), and when viewed in projection with redshift-space distortions, the line-of-sight velocity differences significantly wash out.
As a way to test this method, we varied our line-of-sight velocity cut (which was $\Delta v = \pm 2\,\vthm$), but we did not seen any significant trends to suggest that this method is particularly discriminating.
This echoes the more detailed investigation of \citet{Oman2013}, who showed that, while incorporating line-of-sight velocity can help in estimating time since infall, the reduction in scatter in time since infall at fixed halo-centric distance using velocity information is typically only a few percent.

As a second test, we note that because ejected satellites are significantly stripped of halo mass, they also host fewer satellites in their halo.\footnote{
We thank Mark Fardal for suggesting this test.
}
We find that the average number of satellites within ejected halos is up to $4 \times$ lower than in first-infalling halos of the same stellar mass, especially near massive clusters.
However, while the \textit{relative} reduction of satellite occupation is significant, the \textit{absolute} occupation that is measurable above our mass limit remains low.
For galaxies of $\mstar = 10 ^ {10.1 - 10.5} \msun$, the average number of hosted satellites that are above our mass limit ($10 ^ {9.7} \msun$) drops from 9\% to 3\% for first-infalling versus ejected halos.
Thus, selecting galaxies that host no satellites cannot cleanly demarcate the ejected population.
Conversely, selecting galaxies that do host a satellite can more cleanly select the first-infalling population, but unfortunately this represents only a few percent of all such galaxies, reducing the available statistics significantly.
To test this, we examined the quiescent fraction for galaxies that host a satellite and are within $d / \rthm = 1 - 2.5$ of a more massive host halo, but the resultant uncertainties in the quiescent fraction were to large to discriminate any clear difference.
Incorporating catalogs with fainter galaxies, and thus larger satellite occupations, may help to improve this test, and combining this with line-of-sight velocity information may help to test star formation in ejected versus first-infalling populations more cleanly.
We leave such investigations to future work.

We note that the exact choice of virial radius is somewhat arbitrary, and the complications of ejected satellites could be `defined' away, to some extent, by using a more liberal virial definition to encompass all associated satellites.
However, this alternative has several drawbacks.
Our virial definition of $200\bar{\rho}_{\rm m}$ already is broader than many other commonly used ones, such as $200\rho_{\rm c}$.
Naively extrapolating an NFW density profile, a virial definition whose radius extends to $2.5\,\rthm$---the extent of ejected satellites---would contain an average density of only $\sim 20 \bar{\rho}_{\rm m}$, a density more closely associated with filaments.
A larger virial radius also would encompass more galaxies that are not (yet) physically associated with the host halo and thus have not (yet) likely been affected environmentally.
Thus, the choice of virial radius necessarily represents a trade-off between purity and completeness of the associated satellite population, and a proper treatment of ejected satellites will be important for any reasonable virial definition.

Finally, while our group-finding algorithm assumes that host halos are spherical, their density contours and virial shock fronts can be ellipsoidal, meaning that the physical extent of a halo can extend to more (or less) than a spherical virial radius.
We leave investigation of alternative group finders, in the context of ejected satellites, to future work.

%========================
\subsubsection{Broader implications} \label{sec:implications}

Our results highlight the importance of environmental history in understanding not only star formation but also stellar mass growth in galaxy evolution.
Central galaxies that are really ejected satellites represent a challenge for the standard `halo model' and halo occupation distribution (HOD) formalism \citep[see][for review]{CooraySheth2002}.
Both the SFR/color and stellar mass of ejected satellites do not depend simply on their halo mass, but also on their formation history.
As \fig{m.star_m.halo_rat_v_m.star} showed, ejected satellites have significantly higher stellar mass than their instantaneous halo mass would imply, and they are more likely to be quiescent/red than normal central galaxies of the same stellar mass.
Thus, any modeling approach that seeks to assign stellar mass or SFR/color to halos based only on their instantaneous virial mass, as in a standard halo-model/HOD approach, will underestimate both the stellar mass and the quiescent/red fraction for such central galaxies.
Moreover, because less massive central galaxies are intrinsically less likely to be quiescent/red, both effects bias in the same direction.
This bias will be most important on quasi-linear scales in the vicinity of massive host halos (`1-halo' to `2-halo' transition), where the ejected fraction is 10 - 40\%.
It may be possible to modify the standard halo model to incorporate analytically an associated satellite population that extends out to 2.5 virial radii and spatially overlaps (in a statistical sense) with the population of distinct host halos, though such a modification would require a careful recalibration of the mass function and bias of host halos.
Overall, ejected satellites represent a population for which it is particularly important to follow the formation history of a (sub)halo to understand the galaxy inside.

Relatedly, the biased $\mstar - \mhalo$ relation for ejected satellites, if unaccounted for, could manifest itself as increased scatter in the $\mstar - \mhalo$ relation if averaging over all central galaxies in halo-model/HOD-based analyses.
While this effect is unlikely to be significant at $\mstar \gtrsim 10 ^ {10} \msun$, where the fraction of all central galaxies that are ejected satellites is $< 5\%$ (\fig{ejected.frac_v_m.star}), we expect it to be increasingly significant at lower $\mstar$ where the overall ejected fraction is closer to 10\%.
Note that if 5 - 10\% of galaxies at a given stellar mass have systematically 60\% lower halo mass, this alone translates to a scatter of 0.09 - 0.12 dex in averaging the population.
This represents a non-trivial fraction of empirically measured values of 0.15 - 0.2 dex (see \sect{catalog_simulation}).
Thus, while we do not expect ejected satellites to be the dominant source of scatter in the $\mstar - \mhalo$ relation, they may be a significant component if unaccounted for.

Finally, we note the potential connection of these results to `galactic conformity', that the properties of central galaxies, such as SFR/color, are correlated with those of their satellites \citep{Weinmann2006a}, or more generally, with all neighboring galaxies out to $\sim 4 \mpc$ \citep{Kauffmann2013}.
As \citeauthor{Kauffmann2013} noted, this conformity signal is most pronounced for less massive galaxies that have low SFRs, and can be stronger at larger separations ($> 1 \mpc$).
Ejected satellites provide a natural population to help explain this effect, both because (1) they are relatively low-mass, quiescent central galaxies that reside near more massive halos, which host many quiescent satellite galaxies, and (2) if ejected satellite halos contain their \textit{own} satellites, then both the central and satellite galaxies in such a halo would have passed through a more massive host halo, thus both galaxy types are  more likely to have environmentally quenched (correlated) star formation.
However, \citeauthor{Kauffmann2013} noted that a full explanation of their observed conformity effect would require more than half of all low-mass central galaxies to be ejected satellites, and our results suggest that this fraction is always $< 10\%$ for galaxies with $\mstar > 10 ^ 9 \msun$.
None-the-less, we fully expect that ejected satellites are a population that \textit{should} exhibit some conformity signal and thus would provide some contribution to the overall conformity effect.
We will investigate this issue in future work.

%===================================================================================================
\section*{Acknowledgments}

We thank Michael Blanton, David Hogg, and collaborators for publicly releasing the NYU VAGC, Jarle Brinchmann and the MPA-JHU collaboration for publicly releasing their spectral reductions, and Martin White for simulation data.
We thank Kathryn Johnston and Mark Fardal for enlightening conversations, and Marla Geha for comments on an early draft.
We also thank the anonymous reviewer for many useful comments.
The simulation was analyzed at the National Energy Research Scientific Computing Center.

%===================================================================================================
%\bibliography{biblio}
\bibliography{}

\label{lastpage}

\end{document}